\documentclass[journal=accacs, manuscript=article, layout=twocolumn]{achemso}
\setkeys{acs}{usetitle=true, maxauthors=99}
\usepackage[utf8]{inputenc}
\usepackage{amsmath}
\usepackage{dcolumn}
\usepackage{multirow}
\usepackage[table]{xcolor}
\usepackage[caption=false]{subfig}
\usepackage[export]{adjustbox}
\usepackage{verbatim}
\usepackage{float}
\usepackage{nicefrac}
\mathchardef\mhyphen="2D
\usepackage{tikz}
\newdimen{\myx}
\newdimen{\myy}
\newlength{\myxoff}
\newlength{\myyoff}
\newcommand{\reversiblearrow}[5][1]{
    \path ($(#5) - (#4)$);
    \pgfgetlastxy{\myx}{\myy};
    \pgfpointnormalised{\pgfpoint{\myx}{\myy}};
    \pgfgetlastxy{\myx}{\myy};
    \setlength{\myxoff}{\myy}
    \setlength{\myyoff}{\myx}
    \draw[-left to,color=#2,shorten <=2pt,shorten >=2pt,transform canvas={xshift=-#1\myxoff,yshift=#1\myyoff}] (#4) -- (#5);
    \draw[left to-,color=#3,shorten <=2pt,shorten >=2pt,transform canvas={xshift=#1\myxoff,yshift=-#1\myyoff}] (#4) -- (#5);
    \pgflowlevelsynccm
    }
\usetikzlibrary{shapes,arrows,calc}
\definecolor{ora}{HTML}{e37222}
\definecolor{blu}{HTML}{0065bd}
\tikzset{
    bubbleP/.style={white, fill=white, text opacity=0, ellipse, draw, inner xsep=0pt, inner ysep=1pt, font=\small},
    bubblemediumP/.style={white, fill=white, text opacity=0, ellipse, draw, anchor=center, inner xsep=-3pt, inner ysep=1.5pt, font=\small},
    bubblelargeP/.style={white, fill=white, text opacity=0, ellipse, draw, anchor=center, inner xsep=-5pt, inner ysep=2pt, font=\small},
    bubbleF/.style={black, fill=white, ellipse, draw, inner xsep=0pt, inner ysep=1pt, font=\small},
    bubblemediumF/.style={black, fill=white, ellipse, draw, anchor=center, inner xsep=-3pt, inner ysep=1.5pt, font=\small},
    bubblelargeF/.style={black, fill=white, ellipse, draw, anchor=center, inner xsep=-5pt, inner ysep=2pt, font=\small},
    bubble/.style={black, fill=black, fill opacity=0.2, text opacity=1, ellipse, draw, inner xsep=0pt, inner ysep=1pt, font=\small},
    bubblemedium/.style={black, fill=black, fill opacity=0.2, text opacity=1,  ellipse, draw, anchor=center, inner xsep=-3pt, inner ysep=1.5pt, font=\small},
    bubblelarge/.style={black, fill=black, fill opacity=0.2, text opacity=1, ellipse, draw, anchor=center, inner xsep=-5pt, inner ysep=2pt, font=\small},
    bubbleT/.style={black, fill=blu, fill opacity=0.5, text opacity=1, ellipse, draw, inner xsep=0pt, inner ysep=1pt, font=\small},
    bubblemediumT/.style={black, fill=blu, fill opacity=0.5, text opacity=1,  ellipse, draw, anchor=center, inner xsep=-3pt, inner ysep=1.5pt, font=\small},
    bubblelargeT/.style={black, fill=blu, fill opacity=0.5, text opacity=1, ellipse, draw, anchor=center, inner xsep=-5pt, inner ysep=2pt, font=\small},
    bubbleS/.style={black, fill=ora, fill opacity=0.5, text opacity=1, ellipse, draw, inner xsep=0pt, inner ysep=1pt, font=\small},
    bubblemediumS/.style={black, fill=ora, fill opacity=0.5, text opacity=1, ellipse, draw, anchor=center, inner xsep=-3pt, inner ysep=1.5pt, font=\small},
    bubblelargeS/.style={black, fill=ora, fill opacity=0.5, text opacity=1, ellipse, draw, anchor=center, inner xsep=-5pt, inner ysep=2pt, font=\small},
    }
\newlength{\du}
\newlength{\inv}
\definecolor{black}{rgb}{0.000000, 0.000000, 0.000000}
\usepackage{hyperref}

\title{Selectivity Trends and Role of Adsorbate-Adsorbate Interactions in CO Hydrogenation on Rhodium Catalysts}

\author{Martin Deimel}
\affiliation{Chair for Theoretical Chemistry and Catalysis Research Center, Technical University of Munich, Lichtenbergstr. 4, 85747 Garching, Germany}

\author{Hector Prats}
\affiliation{Department of Chemical Engineering, University College London, Roberts Building, Torrington Place, London WC1E 7JE, UK}

\author{Michael Seibt}
\affiliation{Chair for Theoretical Chemistry and Catalysis Research Center, Technical University of Munich, Lichtenbergstr. 4, 85747 Garching, Germany}

\author{Karsten Reuter}
\affiliation{Fritz-Haber-Institut der Max-Planck-Gesellschaft, Faradayweg 4-6, Berlin, DE 14195, Germany}

\author{Mie Andersen}
\email{mie@phys.au.dk}
\affiliation{Aarhus Institute of Advanced Studies, Aarhus University, Aarhus C, DK‐8000 Denmark}
\alsoaffiliation{Department of Physics and Astronomy - Center for Interstellar Catalysis, Aarhus University, Aarhus C, DK‐8000 Denmark}


\begin{document}

\setlength{\fboxsep}{0pt}

\begin{tocentry}

    \includegraphics[width=3.33in,height=1.87in]{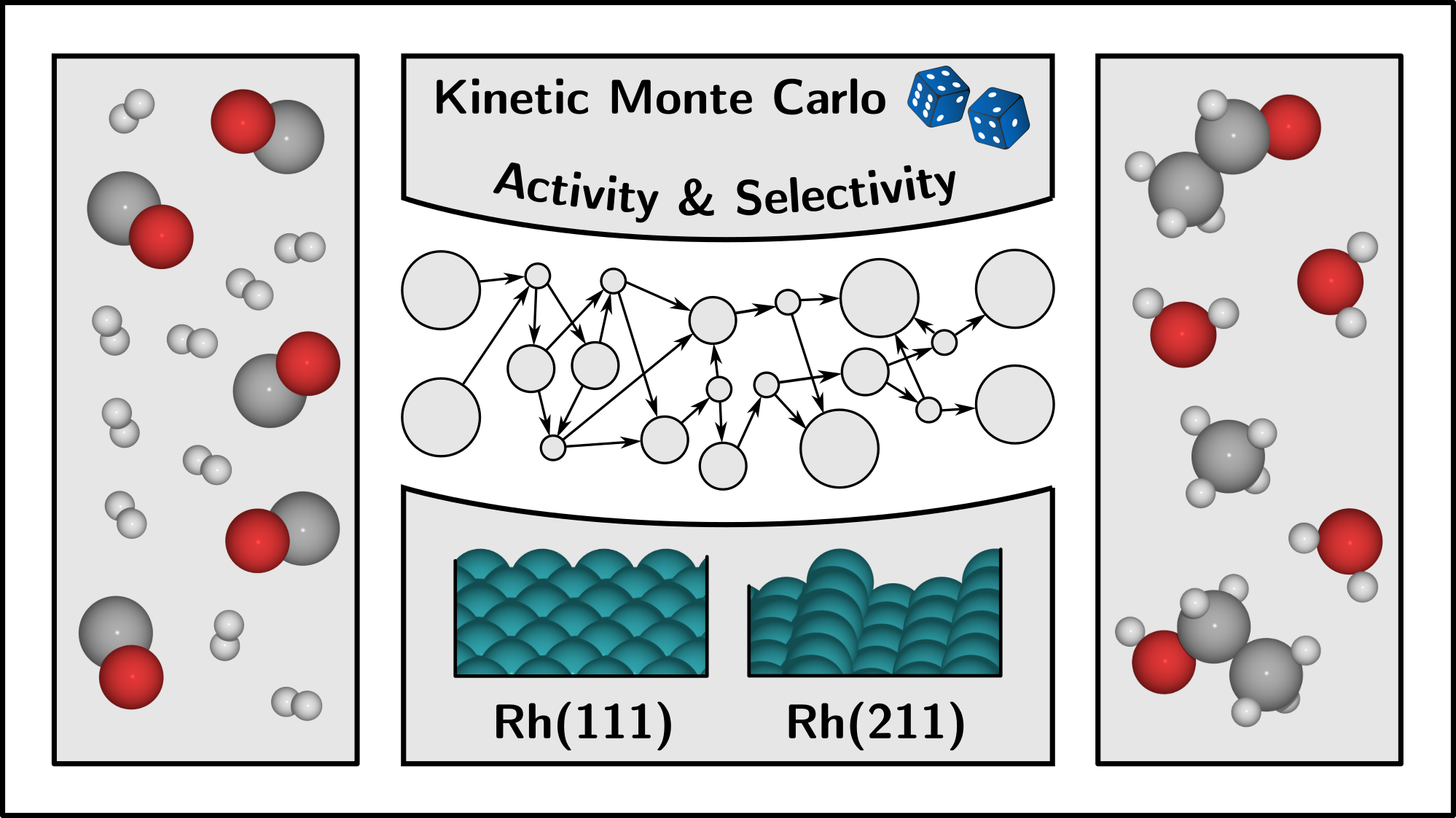}
    
\end{tocentry}

\begin{abstract}

    Predictive-quality computational modeling of heterogeneously catalyzed reactions has emerged as an important tool for the analysis and assessment of activity and activity trends. In contrast, more subtle selectivities and selectivity trends still pose a significant challenge to prevalent microkinetic modeling approaches that typically employ a mean-field approximation (MFA). Here, we focus on CO hydrogenation on Rh catalysts with the possible products methane, acetaldehyde, ethanol and water. This reaction has already been subject to a number of experimental and theoretical studies with conflicting views on the factors controlling activity and selectivity towards the more valuable higher oxygenates. Using accelerated first-principles kinetic Monte Carlo (KMC) simulations and explicitly and systematically accounting for adsorbate-adsorbate interactions through a cluster expansion approach, we model the reaction on the low-index Rh(111) and stepped Rh(211) surfaces. We find that the Rh(111) facet is selective towards methane, while the Rh(211) facet exhibits a similar selectivity towards methane and acetaldehyde. This is consistent with the experimental selectivity observed for larger, predominantly (111)-exposing Rh nanoparticles and resolves the discrepancy to earlier first-principles MFA microkinetic work that found the Rh(111) facet to be selective towards acetaldehyde. While the latter work tried to approximately account for lateral interactions through coverage-dependent rate expressions, our analysis demonstrates that this fails to sufficiently capture concomitant correlations among the adsorbed reaction intermediates that crucially determine the overall selectivity. 
\end{abstract}

\section{Introduction}

    The conversion of syngas (CO and H$_2$) into hydrocarbons and oxygenates is attractive as an alternative source of fuels and chemicals. However, selectivity towards the more useful higher oxygenates such as ethanol and acetaldehyde remains challenging, with methane being a common undesired product.\cite{Luk2017,Ao2018} Many theoretical and experimental works have focused on understanding and tuning especially the selectivity of Rh catalysts, as Rh is generally recognized as one of the most promising elemental catalysts for the direct synthesis of higher oxygenates. It has now become clear that pure Rh catalysts are intrinsically selective towards primarily methane and acetaldehyde,\cite{Yang2016,Schumann2021} whereas ethanol synthesis requires promotors such as Fe or Mn.\cite{Hu2007,Mei2010,Yang2016,Liu2017,Yang2017,Huang2019}
    
    Recent experimental works have suggested that there is an inverse relationship between activity and selectivity for pure Rh catalysts, where an overall higher activity (CO conversion) correlates with a lower selectivity towards acetaldehyde. Explanations offered for this trend are, however, conflicting. Yang~\textit{et~al.}\ have suggested, based on density-functional theory (DFT) and mean-field microkinetics, that it is the nature of the active sites exposed by the catalyst nanoparticles that is the deciding factor, with step sites being highly active and selective towards methane and terrace sites being less active and selective towards acetaldehyde.\cite{Yang2016} This view was recently challenged by Schumann~\textit{et~al.}\cite{Schumann2021} They synthesized Rh nanoparticles of different sizes and found that it is primarily small particles below 2~nm that exhibit high acetaldehyde selectivity and low activity, while the larger particles above 5~nm are the most active and selective towards methane. The surface fraction of edge/corner sites increases for smaller particles and was found to closely follow the selectivity trends. While this could indicate that step or corner sites are the active sites for acetaldehyde synthesis, Schumann~\textit{et~al.}\ proposed instead that the smaller particles support a much higher local CO coverage at both terrace and step sites, which limits the activity due to poisoning, but increases the acetaldehyde selectivity by driving C + CO coupling reactions.
    
    The conflicting views offered in the literature clearly call for more thorough theoretical investigations and improved microkinetic models that can account for the effects of high surface coverages and concomitantly increased lateral adsorbate-adsorbate interactions. In the mean-field study carried out by Yang~\textit{et~al.}\ coverage-dependent rate equations were parametrized and employed to mimic such interactions. However, it is well recognized in the literature that mean-field kinetics cannot properly account for effects of correlations and fluctuations, including fluctuations in the local coverage.\cite{Matera2011,Stamatakis2016,Liu2016,Li2021} A more accurate approach is KMC simulations using a cluster expansion (CE) to treat lateral interactions.\cite{Stamatakis2015,Piccinin2017,Jorgensen2017,Andersen2019KMC,Reuter2016} 
    Unfortunately, this can become very expensive for complex lateral interaction models and large disparities in the time scales of the different processes. In a recent work by Chen~\textit{et~al.}\ on syngas conversion at the Rh(111) surface, some of these challenges were avoided by making the assumption that there is a clear time scale separation between fast diffusion processes and slow reactions such that the system can be solved by alternating between separate KMC and mean-field models.\cite{Chen2018} This assumption may, however, not always be applicable as we will show in this work and as it has also been demonstrated in previous literature works.\cite{Temel2007,Matera2011}
    
    In this work we revisit the question of the role played by step and terrace sites for activity and selectivity trends in syngas conversion over Rh catalysts. We consider the pristine Rh(211) and Rh(111) facets, which are representative of step and terrace sites found at larger nanoparticles where finite-size effects do no longer play a large role. Owing to recent methodological developments in our in-house KMC code kmos\cite{Hoffmann2014} concerning the efficient modeling of lateral interactions and the implementation of an acceleration algorithm\cite{Dybeck2017,Andersen2017akMC} to tackle the time-scale disparity problem, we are able to carry out full-blown KMC simulations employing a CE model for lateral interactions. The results with and without account of lateral interactions are compared to corresponding mean-field kinetics.
    
    The main finding of our work is that -- in contrast to previously parametrized coverage-dependent mean-field models -- KMC simulations that explicitly and systematically account for lateral interactions are able to correctly capture the experimental selectivity trends for large nanoparticles. We show and rationalize why in some cases the lateral interactions have a huge effect on the results (Rh(111) facet), whereas in other cases the effects are negligible (Rh(211) facet). The finding of a breakdown of mean-field kinetics is not restricted to models with lateral interactions. In fact, we show that also in the absence of lateral interactions, reaction-induced inhomogeneities and diffusion limitations can cause the mean-field-predicted activities to deviate substantially from the KMC results (Rh(211) facet). Finally, we show that in order to reach a quantitative agreement with both the selectivity and the activity trends observed in experiments, we need to correct for well-known errors in DFT-predicted adsorption energies such as the persistent overbinding of CO with standard semi-local DFT functionals.\cite{Feibelman2001,AbildPedersen2007,Patra2019}

\section{Methods}

    \subsection{Reaction model and parametrization}
    
\begin{figure*}
            \centering
            \hfill%
            \subfloat{a)\hphantom{\ }\addtocounter{subfigure}{-1}}
            \subfloat{%
            \adjustbox{valign=t}{\resizebox{0.9\columnwidth}{!}
            {
            \begin{tikzpicture}
                \draw [dashed,rounded corners] (5.7\du,\inv-0\du) -- (5.7\du,\inv-2.4\du);
                \draw [dashed,rounded corners] (5.7\du,\inv-3.8\du) -- (5.7\du,\inv-11.75\du) -- (7.7\du,\inv-12.5\du);
                \node[bubbleP] (CO_g)            at (5.7\du,\inv-0\du){CO};
                \node[bubble] (CO_g)            at (5.7\du,\inv-0\du){CO};
                \node[bubbleT] (CO_t)            at (4.3\du,\inv-0.8\du){$\mathrm{CO^{*_{t}}}$};
                \node[bubbleS] (CO_s)            at (6.8\du,\inv-1.2\du){$\mathrm{CO^{*_{s}}}$};
                \node[bubbleF] (C_f)             at (5.7\du,\inv-2.4\du){$\mathrm{C^{*_{f}}}$};
                \node[bubbleF] (CH_f)            at (5.7\du,\inv-3.8\du){$\mathrm{CH^{*_{f}}}$};
                \node[bubbleS] (OH_s)            at (8.2\du,\inv-2.4\du){$\mathrm{OH^{*_{s}}}$};
                \node[bubbleS] (O_s)             at (9.8\du,\inv-3.2\du){$\mathrm{O^{*_{s}}}$};
                \node[bubble] (H2O_g_r)         at (9\du,\inv-1.2\du){$\mathrm{H_2O}$};
                \node[bubblemediumT] (CHO_t)     at (1.8\du,\inv-1\du){$\mathrm{CHO^{*_{t}}}$};
                \node[bubbleT] (O_t)             at (0\du,\inv-1\du){$\mathrm{O^{*_{t}}}$};
                \node[bubblemediumT] (CHOH_t)    at (3.0\du,\inv-2.6\du){$\mathrm{CHOH^{*_{t}}}$};
                \node[bubble] (H2O_g_l)         at (0.8\du,\inv-3.6\du){$\mathrm{H_2O}$};
                \node[bubbleT] (OH_t)            at (0.4\du,\inv-2.3\du){$\mathrm{OH^{*_{t}}}$};
                \node[bubbleT] (CH_t)            at (3.2\du,\inv-3.8\du){$\mathrm{CH^{*_{t}}}$};
                \node[bubbleT] (CH2_t)           at (1.7\du,\inv-4.8\du){$\mathrm{CH_2^{*_{t}}}$};
                \node[bubblemediumT] (CHCO_t)    at (4.2\du,\inv-5.5\du){$\mathrm{CHCO^{*_{t}}}$};
                \node[bubbleS] (CH2_s)           at (9.7\du,\inv-4.8\du){$\mathrm{CH_2^{*_{s}}}$};
                \node[bubblemediumS] (CHCO_s)    at (7.2\du,\inv-5.5\du){$\mathrm{CHCO^{*_{s}}}$};
                \node[bubblemediumT] (CH3CO_t)   at (4.2\du,\inv-8.3\du){$\mathrm{CH_3CO^{*_{t}}}$};
                \node[bubblemediumT] (CH2CO_t)   at (4.2\du,\inv-6.9\du){$\mathrm{CH_2CO^{*_{t}}}$};
                \node[bubblelargeT] (CH3CHO_t)   at (2.5\du,\inv-9.6\du){$\mathrm{CH_3CHO^{*_{t}}}$};
                \node[bubblelargeP] (CH3CHO_g)   at (5.7\du,\inv-9.6\du){$\mathrm{CH_3CHO}$};
                \node[bubblelarge] (CH3CHO_g)   at (5.7\du,\inv-9.6\du){$\mathrm{CH_3CHO}$};
                \node[bubblelargeT] (CH3CHOH_t)  at (3.7\du,\inv-11\du){$\mathrm{CH_3CHOH^{*_{t}}}$};
                \node[bubble] (CH4_g_l)         at (1.7\du,\inv-7.9\du){$\mathrm{CH_{4}}$};
                \node[bubbleT] (CH3_t)           at (1.7\du,\inv-6.4\du){$\mathrm{CH_3^{*_{t}}}$};
                \node[bubble] (CH4_g_r)         at (9.7\du,\inv-7.9\du){$\mathrm{CH_{4}}$};
                \node[bubbleS] (CH3_s)           at (9.7\du,\inv-6.4\du){$\mathrm{CH_3^{*_{s}}}$};
                \node[bubblemediumS] (CH3CO_s)   at (7.2\du,\inv-8.3\du){$\mathrm{CH_3CO^{*_{s}}}$};
                \node[bubblemediumS] (CH2CO_s)   at (7.2\du,\inv-6.9\du){$\mathrm{CH_2CO^{*_{s}}}$};
                \node[bubblelargeS] (CH3CHO_s)   at (8.9\du,\inv-9.6\du){$\mathrm{CH_3CHO^{*_{s}}}$};
                \node[bubblelargeT] (CH3CH2OH_t) at (3.7\du,\inv-12.5\du){$\mathrm{CH_3CH_2OH^{*_{t}}}$};
                \node[bubblelargeP] (CH3CH2OH_g) at (7.7\du,\inv-12.5\du){$\mathrm{CH_3CH_2OH}$};
                \node[bubblelarge] (CH3CH2OH_g) at (7.7\du,\inv-12.5\du){$\mathrm{CH_3CH_2OH}$};
                \node[bubblelargeS] (CH3CHOH_s)  at (7.7\du,\inv-11\du){$\mathrm{CH_3CHOH^{*_{s}}}$};
                \pgflowlevelsynccm
                \foreach \from/\to in {CO_g/CO_t,CO_g/CO_s,CO_t/CHO_t,CHO_t/CHOH_t,CHOH_t/CH_t,CHOH_t/OH_t,CO_t/C_f,CO_t/OH_t,OH_t/O_t,OH_t/H2O_g_l,CO_s/C_f,CO_s/OH_t,OH_s/O_s,OH_s/H2O_g_r,C_f/CH_f,CH_f/CH2_t,CH_t/CH2_t,CH2_t/CH3_t,CH3_t/CH4_g_l,CH_f/CH2_s,CH2_s/CH3_s,CH3_s/CH4_g_r,CH3_s/CH4_g_r,CH_t/CHCO_t,CO_t/CHCO_t,CHCO_t/CH2CO_t,CH2CO_t/CH3CO_t,CH3CO_t/CH3CHO_t,CH3CHO_t/CH3CHOH_t,CH3CHOH_t/CH3CH2OH_t,CH_f/CHCO_s,CO_s/CHCO_s,CHCO_s/CH2CO_s,CH2CO_s/CH3CO_s,CH3CO_s/CH3CHO_s,CH3CHO_s/CH3CHOH_s,CH3CHOH_s/CH3CH2OH_g,CH3CH2OH_t/CH3CH2OH_g,CH3CHO_t/CH3CHO_g,CH3CHO_s/CH3CHO_g}
                    \reversiblearrow[1]{black}{black}{\from}{\to};
                \pgflowlevelsynccm
            \end{tikzpicture}
            }\label{fig:network_211}}%
            }\hfill%
            \subfloat{b)\hphantom{\ }\addtocounter{subfigure}{-1}}
            \subfloat{%
            \adjustbox{valign=t}{\resizebox{0.79\columnwidth}{!}
            {
            \begin{tikzpicture}
                \node[bubble] (CO_g)            at (5.7\du,\inv-0\du){CO};
                \node[bubbleT] (CO_t)            at (4.3\du,\inv-0.8\du){$\mathrm{CO^{*_{t}}}$};
                \node[bubblemediumT] (CHO_t)     at (1.8\du,\inv-1\du){$\mathrm{CHO^{*_{t}}}$};
                \node[bubblemediumT] (CHOH_t)    at (3.0\du,\inv-2.6\du){$\mathrm{CHOH^{*_{t}}}$};
                \node[bubble] (H2O_g_l)         at (0.8\du,\inv-3.6\du){$\mathrm{H_2O}$};
                \node[bubbleT] (OH_t)            at (0.4\du,\inv-2.3\du){$\mathrm{OH^{*_{t}}}$};
                \node[bubbleT] (CH_t)            at (3.2\du,\inv-3.8\du){$\mathrm{CH^{*_{t}}}$};
                \node[bubbleT] (CH2_t)           at (1.7\du,\inv-4.8\du){$\mathrm{CH_2^{*_{t}}}$};
                \node[bubblemediumT] (CHCO_t)    at (4.2\du,\inv-5.5\du){$\mathrm{CHCO^{*_{t}}}$};
                \node[bubblemediumT] (CH3CO_t)   at (4.2\du,\inv-8.3\du){$\mathrm{CH_3CO^{*_{t}}}$};
                \node[bubblemediumT] (CH2CO_t)   at (4.2\du,\inv-6.9\du){$\mathrm{CH_2CO^{*_{t}}}$};
                \node[bubblelargeT] (CH3CHO_t)   at (2.5\du,\inv-9.6\du){$\mathrm{CH_3CHO^{*_{t}}}$};
                \node[bubblelarge] (CH3CHO_g)   at (5.7\du,\inv-9.6\du){$\mathrm{CH_3CHO}$};
                \node[bubblelargeT] (CH3CHOH_t)  at (3.7\du,\inv-11\du){$\mathrm{CH_3CHOH^{*_{t}}}$};
                \node[bubble] (CH4_g_l)         at (1.7\du,\inv-7.9\du){$\mathrm{CH_{4}}$};
                \node[bubbleT] (CH3_t)           at (1.7\du,\inv-6.4\du){$\mathrm{CH_3^{*_{t}}}$};
                \node[bubblelargeT] (CH3CH2OH_t) at (3.7\du,\inv-12.5\du){$\mathrm{CH_3CH_2OH^{*_{t}}}$};
                \node[bubblelarge] (CH3CH2OH_g) at (7.7\du,\inv-12.5\du){$\mathrm{CH_3CH_2OH}$};
                \pgflowlevelsynccm
                \foreach \from/\to in {CO_g/CO_t,CO_t/CHO_t,CHO_t/CHOH_t,CHOH_t/CH_t,CHOH_t/OH_t,
                OH_t/H2O_g_l,CH_t/CH2_t,CH2_t/CH3_t,CH3_t/CH4_g_l,CH_t/CHCO_t,CO_t/CHCO_t,CHCO_t/CH2CO_t,CH2CO_t/CH3CO_t,CH3CO_t/CH3CHO_t,CH3CHO_t/CH3CHOH_t,CH3CHOH_t/CH3CH2OH_t,CH3CH2OH_t/CH3CH2OH_g,CH3CHO_t/CH3CHO_g}
                    \reversiblearrow[1]{black}{black}{\from}{\to};
                \pgflowlevelsynccm
            \end{tikzpicture}
            }\label{fig:network_111}}%
            }\hfill%
            \phantom{}%
            \vskip0.1cm%
            \subfloat{c)\hphantom{\ }\addtocounter{subfigure}{-1}}%
            \subfloat{%
            \adjustbox{valign=t}{%
            \fbox{\includegraphics[width=0.3\textwidth]{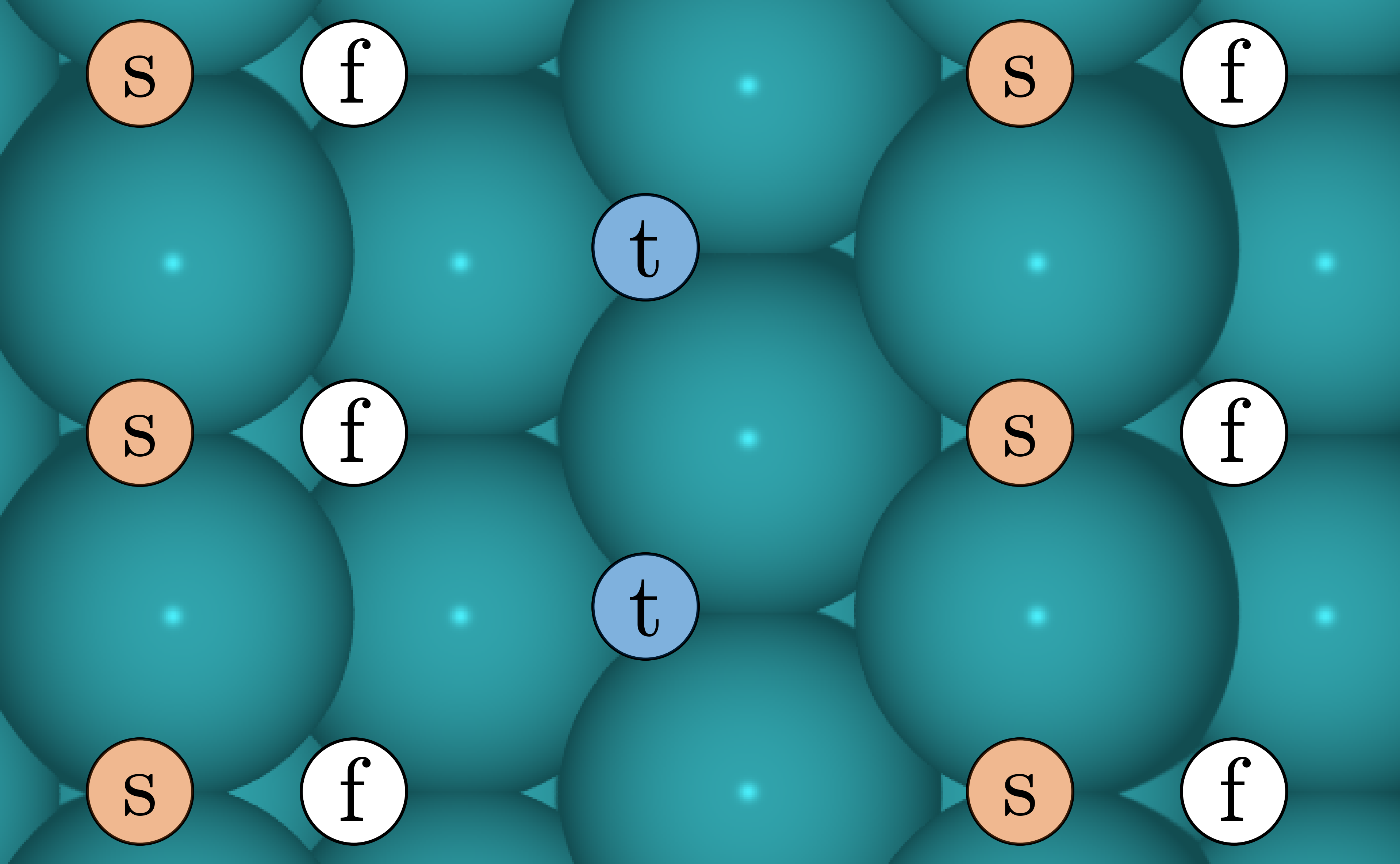}}%
            \label{fig:211_facet}%
            }%
            }%
            \hspace{2cm}
            \subfloat{d)\hphantom{\ }\addtocounter{subfigure}{-1}}%
            \subfloat{%
            \adjustbox{valign=t}{%
            \fbox{\includegraphics[width=0.3\textwidth]{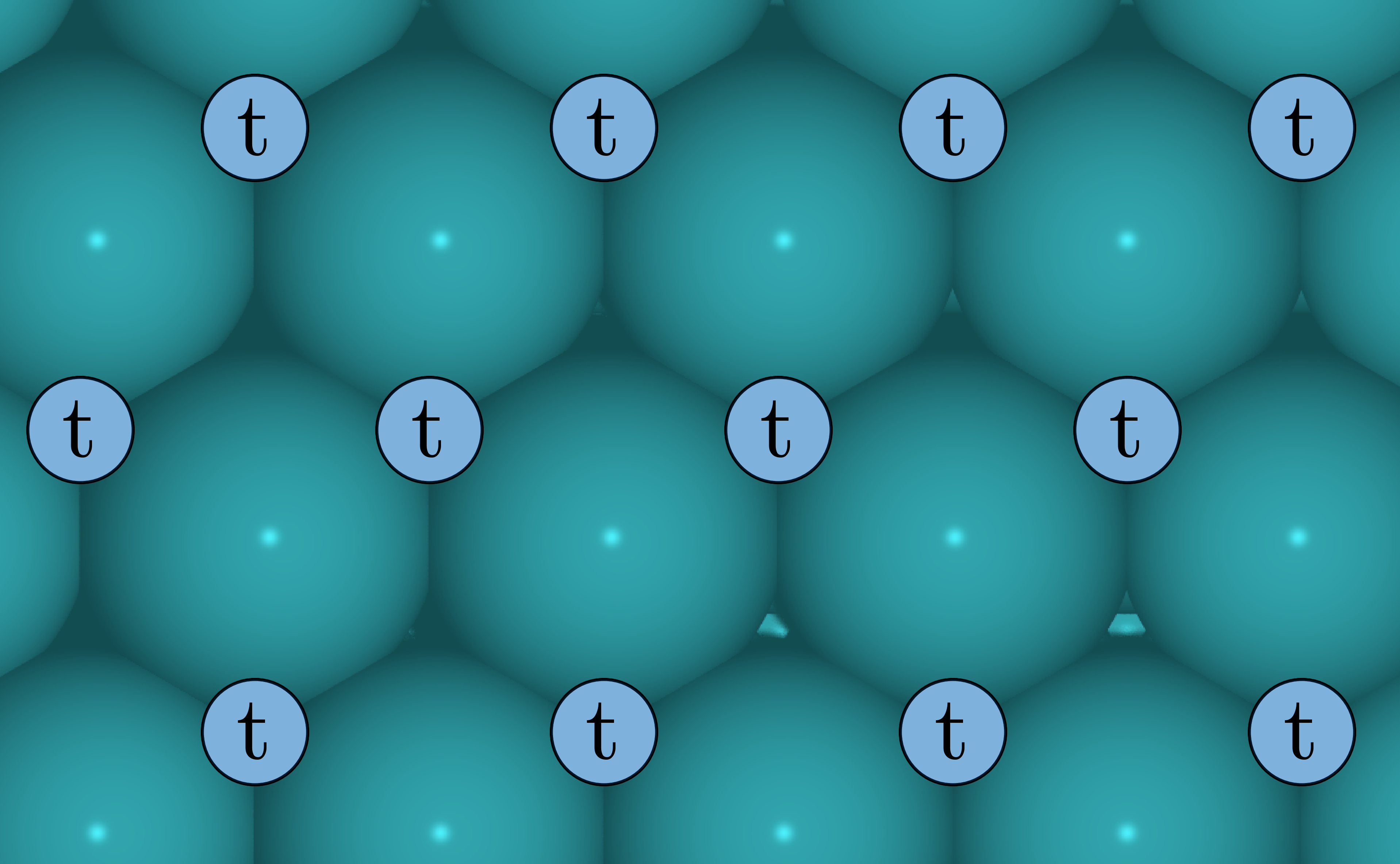}}%
            \label{fig:111_facet}%
            }%
            }
            \caption{Reaction networks of (a) the Rh(211) and (b) the Rh(111) KMC models, excluding diffusion steps (these connect e.g.\ $\mathrm{OH^{*_t}}$ and $\mathrm{OH^{*_s}}$). For hydrogenation reactions an additional $\mathrm{H^{*_{S/T}}}$ species is implied. On the right side of the dashed line in (a) are shown the elementary steps occurring on the $s$ sites (orange) and on the left side the corresponding reactions on the $t$ sites (blue). Adsorbates on $f$ sites are shown in white and gaseous molecules are shown in gray. Images of the corresponding Rh(211) and Rh(111) facets with the respective sites are shown in (c) and (d).}
            \label{fig:facets}
        \end{figure*}
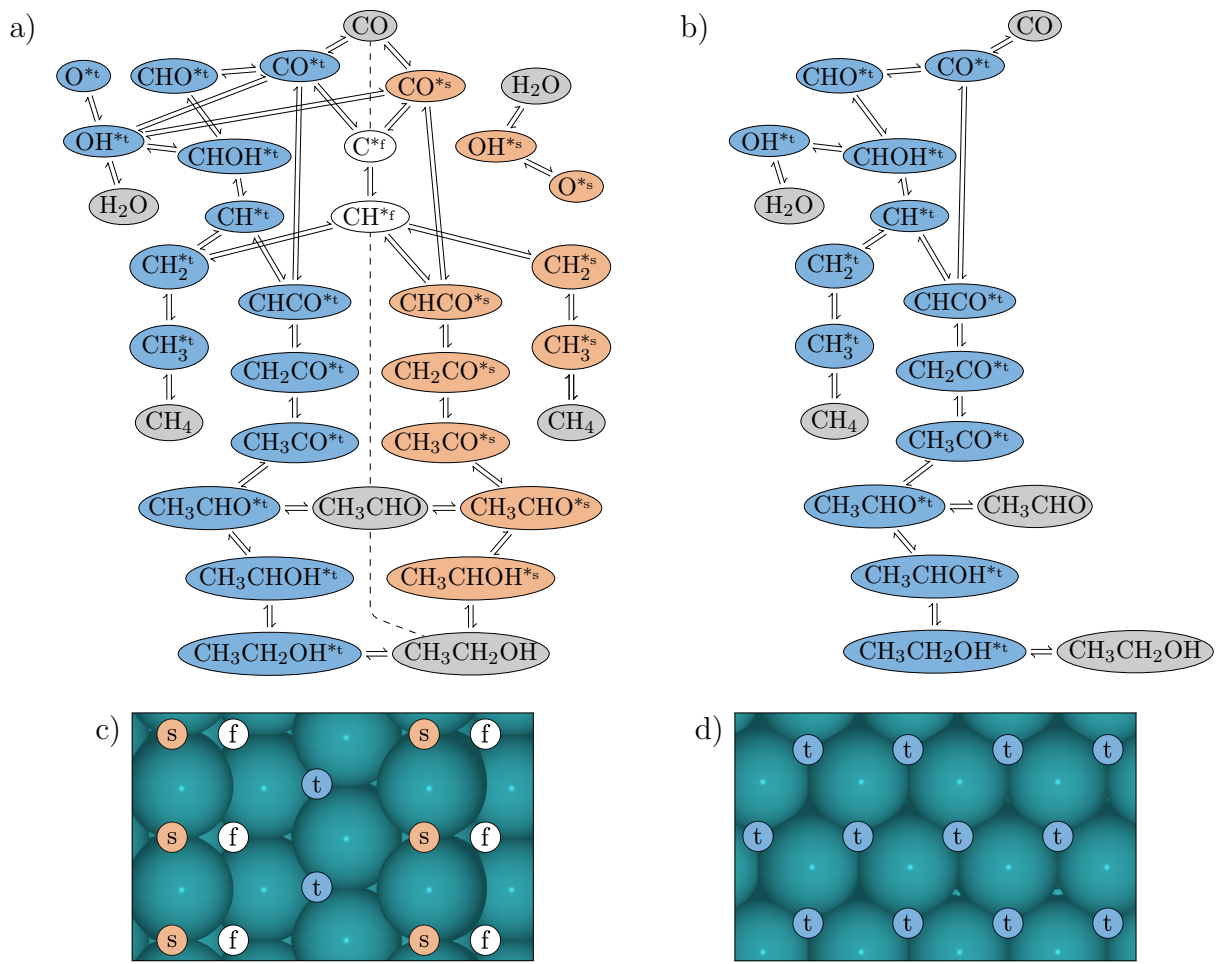

        The reaction networks employed for the Rh(211) and Rh(111) facets are shown in Figure~\ref{fig:facets} and were inspired by the work of Yang~\textit{et~al.}\cite{Yang2016} As in our previous works on CO hydrogenation over stepped metal surfaces,\cite{Andersen2017akMC,Deimel2020} and extending over the work of Yang~\textit{et~al.}, we use a highly resolved active site representation for the Rh(211) facet consisting of a terrace site $t$, an upper step site $s$ and a lower four-fold coordinated step site $f$. Only the two latter sites were considered by Yang~\textit{et~al.} The full reaction networks we employed can be found in Section~S1 of the Supplementary Information (SI). DFT data calculated at both low and high CO coverage were taken from Yang~\textit{et~al.}\ and used to parametrize a CE model, which is used in connection with the KMC simulations. Some additional DFT calculations were carried out by us using the Quantum Espresso code \cite{Giannozzi2009QE} with the BEEF-vdW functional \cite{Wellendorff2012BEEF} and the exact same numerical settings as used by Yang~\textit{et~al.}, see Section~S2 in the SI.

        Adsorption energies are referenced to CO, $\mathrm{H_2O}$ and $\mathrm{CH_3OH}$ using the formation energy approach
        \[ E^x_{\mathrm{form}} = E^x_{\mathrm{slab+ads}} - E_{\mathrm{slab}} - \sum_{i \in x} n^x_i \mu_i \quad , \]
        with the formation energy $E^x_{\mathrm{form}}$ of adsorbate $x$, the total energy of the adsorbed species on the slab $E^x_{\mathrm{slab+ads}}$, that of the clean slab $E_{\mathrm{slab}}$, and the gas phase reference energy $\mu_i$ of atom $i$ as well as the occurrence $n^x_i$ of atom $i$ in the adsorbate $x$. Adsorbate-adsorbate interactions are incorporated using a CE which is terminated after the first term corresponding to pairwise nearest neighbor interactions
        \[ E^M_{\mathrm{form}} = \sum_{x \in M} n^M_x E^x_{\mathrm{form}} + \sum_{j} k^M_j \varepsilon_j \quad , \]
        with the occurrence $n^M_x$ and formation energies $E^x_{\mathrm{form}}$ of the individual adsorbates $x$ in structure $M$ and the occurrence $k^M_j$ and interaction energy $\varepsilon_j$ of the pairwise interaction $j$. The interaction energies (cf.\ Table~S2 in the SI) are obtained by solving a set of linear equations for a number of different structures $M$. As the surface at the investigated conditions is mostly covered with CO, we consider only self-interactions between the CO molecules, interactions between CO and the remaining adsorbates (except H), as well as selected other interactions (e.g.\ between C and CH at the $f$ site) following Yang~\textit{et~al.} All other interactions are neglected due to their low probability of occurrence. Following common practice,\cite{Lausche2013,Yang2016} H is adsorbed at special "hydrogen reservoir" sites, which reflects the assumption that it can intercalate into sublattice sites because of its smaller size compared to the other adsorbates. In the KMC models, we implement this by treating H in a mean-field ansatz, following our previous work.\cite{Andersen2017akMC} This means that it is not an actual species but only represented via an effective H coverage. We use different energetics (and thereby coverage) depending on whether the H atom is involved in reactions taking place at an $s$ or a $t$ site (for H we denote these reservoir sites as $S$ and $T$ sites both in the KMC and mean-field models). Our treatment thereby implicitly assumes that there are no spatial correlations in the distribution of H at the surface, which should be a good approximation when taking into account that H diffusion and adsorption/desorption are fast processes. Note, however, that the assumption that H does not block any surface sites, nor interacts with the other adsorbates, would most likely break down under high-pressure reaction conditions.
        
        For the rate constants of elementary steps corresponding to adsorption, desorption, reaction, and diffusion, we used standard expressions including zero-point energies and other enthalpy/entropy corrections within the harmonic approximation (adsorbates and transition states) or the ideal gas approximation (gas species) from the thermochemistry module of the Atomic Simulation Environment software package,\cite{ase-paper-1,ase-paper-2} cf.\ our previous work \cite{Andersen2017akMC}. Vibrational frequencies are taken from Yang~\textit{et~al.}\cite{Yang2016} In our previous work,\cite{Andersen2017akMC} diffusion steps of CO, O, OH, CH, CH$_2$, and CH$_3$ were found not to be rate-limiting and therefore we stick to approximate barriers calculated for Re(0001) from Hahn~\textit{et~al.}\cite{Hahn2014} For same-site diffusion of C at $f$ sites and CH at $t$ sites we use values calculated for Rh(211) from our previous work.\cite{Andersen2017akMC} Diffusion of all other species is neglected. To allow for comparison of our simulations to those of Yang~\textit{et~al.}\cite{Yang2016}, we use the same reaction conditions of $p_{\mathrm{CO}}$ = 13.33~bar, $p_{\mathrm{H_2}}$ = 6.66~bar, and $p_{\mathrm{H_2O}}$ = $p_{\mathrm{CH_4}}$ = $p_{\mathrm{CH_3CHO}}$ = $p_{\mathrm{CH_3CH_2OH}}$ = 0~bar at the three different temperatures 523~K, 585~K, and 650~K.
        
        In order to model the effect of lateral interactions on the reaction kinetics during the simulations, the energy barriers $E^a$ of the elementary steps for the possible lattice configurations are linearly approximated using Br{\o}nsted-Evans-Polanyi (BEP) relations \cite{Bronsted1928BEP,Evans1937BEP}
        \[ E^a = \alpha \left( \Delta E_{\mathrm{FS}} - \Delta E_{\mathrm{IS}} \right) + E^a_0 \quad , \]
        with the energy shift $\Delta E_{\mathrm{FS}}$ ($\Delta E_{\mathrm{IS}}$) of the final (initial) state due to lateral interactions, the zero-coverage barrier $E^a_0$, and the parameter $\alpha$ with values in the interval $[0,1]$ representing a reactant like (0) or product like (1) transition state. DFT-calculated energetics for determining the $\alpha$ parameters shown in Table~S3 are taken from Yang~\textit{et~al.}\ and Andersen~\textit{et~al.}\cite{Yang2016, Andersen2017akMC} Although BEP relations entail some energetic uncertainty, the magnitude is on the order of the error introduced by semi-local DFT.\cite{Stamatakis2015} The approach of combining CE with BEP relations is used across different KMC frameworks and effectively reduces the computational burden for the inclusion of lateral interactions. Previous studies employing this approach were able to quantitatively capture experimental observations without the need to explicitly calculate all possible lattice configurations, which is especially critical with an increasing number of sites and adsorbates in the reaction network under consideration.\cite{Wu2012,Nielsen2013,Piccinin2014,Stamatakis2015}

    \subsection{Kinetic Monte Carlo}
    
        The KMC simulation technique allows for a numerical solution to the time evolution of the spatial distribution of the adsorbates on the coarse-grained sites of the catalytic surface, along with related properties such as the occurrence of individual reaction pathways and the catalytic activity and selectivity in terms of the turn-over-frequencies (TOFs) for the formation of the reaction products. Since individual elementary steps are executed step-by-step with probabilities reflected by their rate constants, KMC can suffer from a performance bottleneck in case of processes that occur on disparate timescales, e.g.\ fast diffusion and slow reactions. This timescale disparity challenge is here tackled through an acceleration algorithm developed by Dybeck~\textit{et~al.}\cite{Dybeck2017} In a recent work \cite{Andersen2017akMC} we implemented this algorithm in the kmos code\cite{Hoffmann2014} and used it to study trends in CO methanation activity over stepped transition metals. While some challenging cases where observed, the algorithm was found to work well for the mainly CO-covered Rh(211) facet. Our previous work disregarded lateral interactions, however, and these can play an important role on the outcome of a reaction as shown in this work. Both repulsive and attractive interactions with neighboring adsorbates can alter the energetics of elementary steps and lead to changes in their individual rates. In this work we used the acceleration algorithm in connection with the recently developed on-the-fly backend in kmos \cite{Lorenzi2017Thesis,Seibt2018}. This backend features increased performance and reduced memory requirements for models with many lateral interactions compared to the original backend, since the rates of processes affected by lateral interactions are calculated at runtime according to the above-mentioned CE model and BEP relations.
        
        The error bars for the TOFs are obtained from a Bayesian error analysis. Reactions in KMC follow the Poisson distribution
        \[P\left(n|v\right) = \frac{e^{-v}v^{n}}{n!} \quad \text{,}\]
        where $P\left(n|v\right)$ is the probability for observing $n$ turnovers during a fixed simulation time $t$, given the expected value $v$. Here, we are rather interested in calculating the probability distribution for $v$, given a (possibly small or even zero) number of observed turnover events $n$. This posterior distribution can be obtained via Bayes theorem
        \begin{align*}
        P\left(v|n\right) &= \frac{P\left(n|v\right)P_0\left(v\right)}{\int_0^{\infty}P\left(n|v\right)P_0\left(v\right) dv}\\ &= \frac{e^{-v}v^{n}P_0\left(v\right)}{\int_0^{\infty}e^{-v}v^{n}P_0\left(v\right) dv}
        \end{align*}
        with the prior probability distribution $P_0\left(v\right)$, which we assume to be a constant $C$ as there is no \textit{a priori} information on the TOF. To normalize $P_0\left(v\right)$ we set a $v_{\text{max}}$, i.e.\ $C=1/v_{\text{max}}$, to a value high enough that the following approximation holds
        \[\frac{e^{-v}v^{n}}{\int_0^{v_{\text{max}}}e^{-v}v^{n}dv} \approx \frac{e^{-v}v^{n}}{\int_0^{\infty}e^{-v}v^{n}dv} = \frac{e^{-v}v^{n}}{n!}\text{.}\]
        This gives the posterior probability distribution
        \[P\left(v|n\right) = \frac{e^{-v}v^{n}}{n!}\text{,}\]
        which is also a Poisson distribution. Note that this also gives us the probability distribution for the TOF, since the TOF is simply $v$ divided by the simulation time $t$. The most likely value of the TOF is determined from the maximum of the posterior distribution, and the upper and lower bounds of the error bars are obtained as the smallest credible interval that contains 99\% of the total probability mass.
    
        Sensitivity analysis was carried out by calculating the degree of rate control (DRC) proposed by Campbell and coworkers \cite{Stegelmann2009DRC} for each TS $i$ ($X_{\mathrm{RC},i}$)
        \[ X_{\mathrm{RC},i} = \left( \frac{\partial \ln r}{\partial \left( \frac{-G_{i}}{RT} \right)} \right)_{G_{j \neq i}} \]
        with the rate $r$, the free energy $G_i$ of TS $i$, the universal gas constant $R$ and the absolute temperature $T$. In KMC the derivative was approximated by the finite-difference expression
        $X_{\mathrm{RC},i} = \left( \frac{\ln r_{+} - \ln r_{-}}{\frac{-0.2 \:{\rm eV}}{RT}} \right)_{G_{j \neq i}}$ with $r_{+}$ ($r_{-}$) being the rate for an increase (decrease) of the TS energy by 0.1~eV. For the determination of the associated error bars, the respective upper $v_+$ and lower bounds $v_-$ of the 99\% credible interval for the two energy modifications are combined to contain $v_+$ of one and $v_-$ of the other simulation and \textit{vice versa}. The upper (lower) limit of the selectivities of methane, acetaldehyde, and ethanol are obtained by considering $v_+$ ($v_-$) of the 99\% credible interval of one product and the respective $v_-$ ($v_+$) of the other two products.
    
        The pair probabilities of second order processes are obtained by storing the time-integrated counts of lattice configurations in which the process can be executed over the entire simulation, and the counts are then normalized by the total simulation time and the size of the system. All simulations were run for $5\times10^{7}$ steps to reach a steady-state and subsequently until 11 (Rh(111)) or 26 turnovers (Rh(211)) of acetaldehyde were observed or the total KMC simulation time reached one week.
        The simulations were carried out using a (10$\times$10) lattice for Rh(211) and a (25$\times$25) lattice for Rh(111) with periodic boundary conditions. Further details about the KMC simulation settings and the convergence tests carried out for both the lattice size as well as the parameters employed in the acceleration algorithm can be found in Section S3 of the SI.
       
    \subsection{Mean-field microkinetic modeling}
    
        For comparison to the KMC simulations, we also carried out microkinetic simulations using the CatMAP software package.\cite{Medford2015CatMAP} CatMAP employs the MFA, meaning that the spatial distribution of the adsorbates is further coarse-grained into a mean coverage of each site type, thereby also neglecting coverage fluctuations around the mean. This results in a set of coupled rate equations, which are solved at steady state. All of our MFA simulations were performed without lateral interactions, as the parametrization of coverage-dependent rate equations is not a topic of this work. We emphasize that the MFA simulations used exactly the same reaction network and adsorption energetics as our KMC simulations, so that differences between both simulation approaches are entirely due to the MFA employed in the prior. 

\section{Results and discussion}

\begin{figure*}
        \centering
        \hfill%
        \subfloat{a)\addtocounter{subfigure}{-1}}
        \subfloat{%
        \includegraphics[valign=t,width=0.45\textwidth]{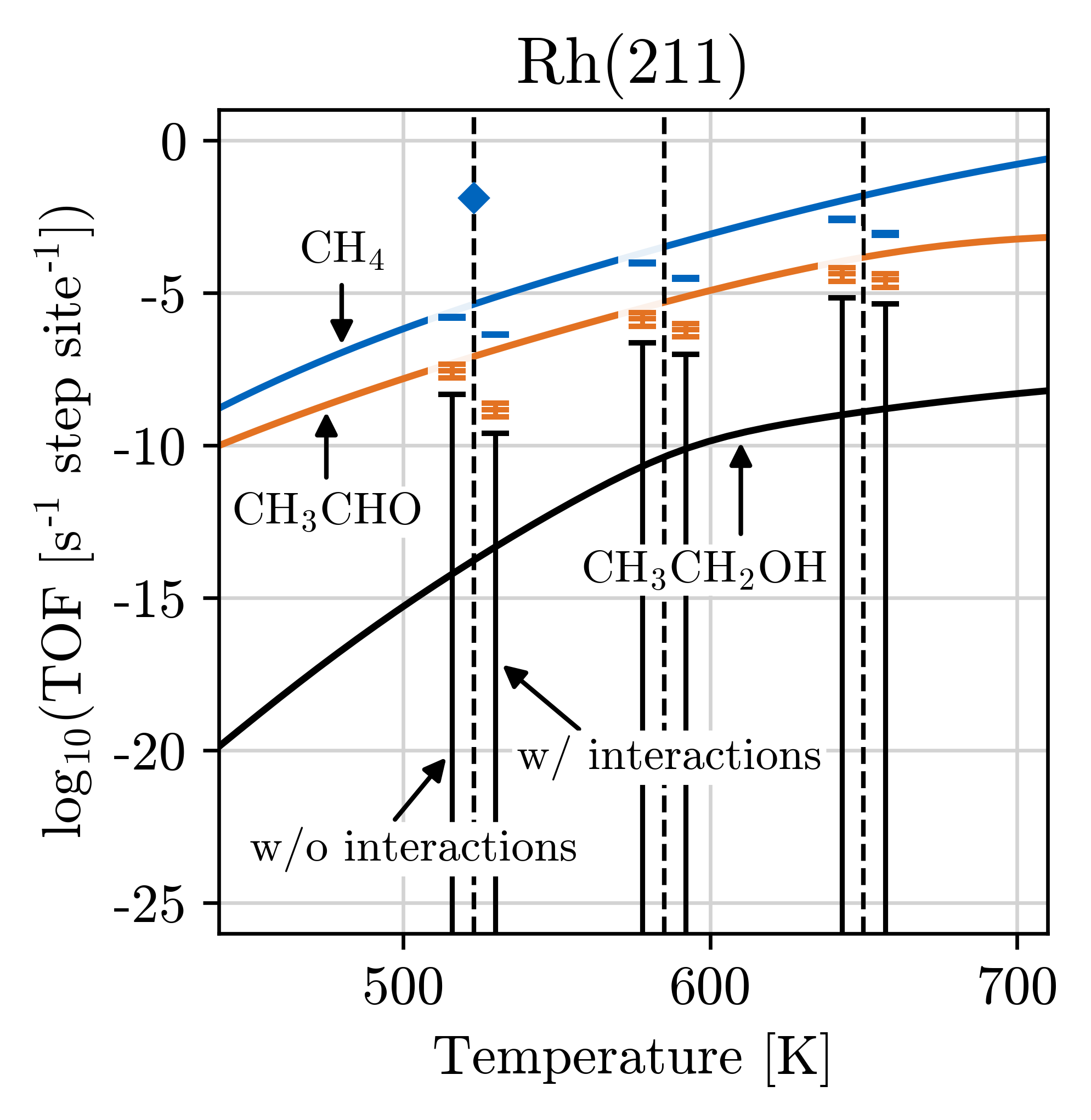}\label{fig:TOF_comparison_211}}%
        \hfill%
        \subfloat{b)\addtocounter{subfigure}{-1}}
        \subfloat{%
        \includegraphics[valign=t,width=0.45\textwidth]{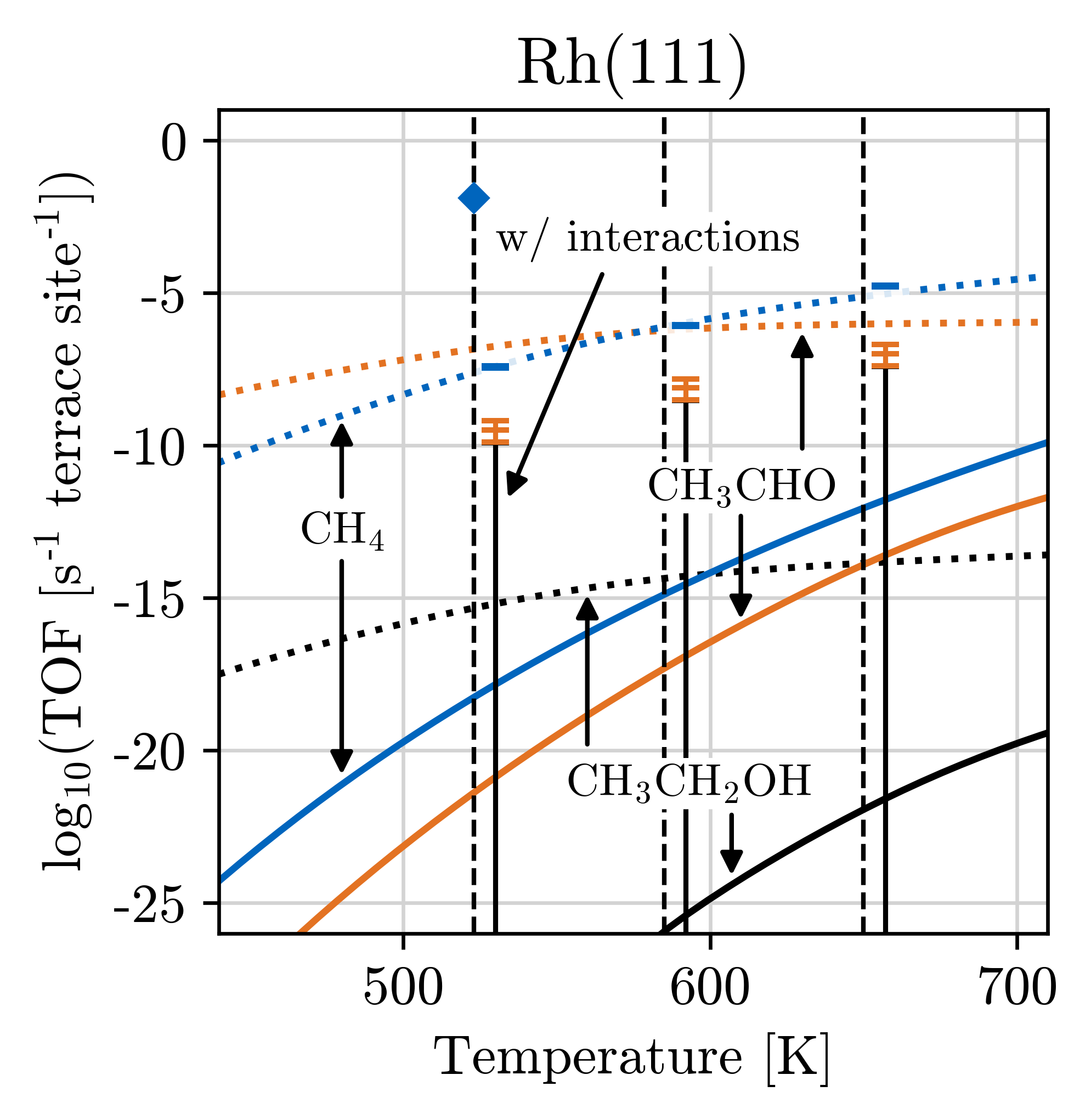}\label{fig:TOF_comparison_111}}%
        \hfill\phantom{}%
        \caption{TOFs as a function of the temperature for (a) the Rh(211) and (b) the Rh(111) facet for CH$_4$ (blue), CH$_3$CHO (orange), and CH$_3$CH$_2$OH (black). The MFA results without (with) lateral interactions are shown with solid lines (dotted lines) and the KMC results are shown at 523~K, 585~K, and 650~K (dashed vertical lines) without (left offset) and with (right offset) lateral interactions. The blue diamond at 523~K is the experimentally measured total TOF (primarily methane) of the largest nanoparticles with diameters above 5~nm from Schumann~\textit{et~al.}\cite{Schumann2021}.
        }
    \end{figure*}

    We begin by presenting MFA and KMC results for the Rh(211) facet in Figure~\ref{fig:TOF_comparison_211}. The KMC results are reported at three different temperatures without (left offset) and with (right offset) lateral interactions. Without lateral interactions, a comparison of our MFA results (solid lines) and KMC results (left offset) reveals that the activities are slightly overestimated in the MFA model by about a factor of 2.7-6.0 and that both models show similar selectivity trends towards methane rather than acetaldehyde. Ethanol turnovers are not observed in the KMC simulations - hence the large error bars - which is consistent with the very low TOFs obtained in the MFA model. We will come back to the reason for the differences in actual TOF prediction between MFA and KMC below. 

    Turning to the effect of lateral interactions, we can observe from the KMC results that they do not influence the TOFs much. Our MFA results are only presented without lateral interactions as the parametrization of these are not available for our modified active site representation. However, in the original Rh(211) MFA model from Yang~\textit{et~al.} (see Figure~S1 in the SI), lateral interactions do influence the TOFs, especially at lower temperatures. At 523~K the difference ranges from a factor of about 300 for ethanol to about 4000 for acetaldehyde.
    
    Before diving into a deeper analysis of the differences between the different models for the Rh(211) facet, we present also the TOFs for the Rh(111) facet, see Figure~\ref{fig:TOF_comparison_111}. This facet contains only a terrace site, and thus the active site representation and reaction network we employ (cf.\ Section~S1.3 in the SI) is in this case completely identical to that of Yang~\textit{et~al.} Comparing the MFA results with and without interactions (dotted and solid lines) we can see that for this facet the influence of the lateral interactions is huge -- at all temperatures the catalyst is essentially inactive if lateral interactions are not taken into account. Furthermore, it is surprising to note that the lateral interactions in the MFA model, which were parametrized using a second-order expansion in the coverage, give rise to a change in the selectivity trends. That is, without interactions the catalyst is always selective towards methane, but with interactions the catalyst is selective towards acetaldehyde at the experimentally relevant temperature below about 600 K. We recall that the MFA results for Rh(211) and Rh(111) with interactions were used by Yang~\textit{et~al.} to propose that the activity-selectivity trends obtained with experimentally synthesized Rh nanoparticles are caused by varying amounts of step and terrace sites at these particles, where step (terrace) sites would then be the active sites for methane (acetaldehyde) formation, respectively.\cite{Yang2016}
    
    However, as discussed in the introduction, this explanation was recently challenged by Schumann~\textit{et~al.}\cite{Schumann2021}\ based on new and more detailed experiments that showed that large particles above 5~nm, which are expected to predominantly expose the Rh(111) facet, are actually selective towards methane and not acetaldehyde. Interestingly, we can fully confirm this from our KMC simulations with lateral interactions, where at all simulated temperatures we find that the Rh(111) facet is indeed selective towards methane, see Figure~\ref{fig:TOF_comparison_111}. Without interactions we do not observe any meaningful TOFs for any of the products (not shown), which is consistent with the very low TOFs obtained in the MFA model without interactions. Thus, our KMC results confirm that lateral interactions are of paramount importance for the Rh(111) facet, but the CE used to parametrize these interactions in KMC gives qualitatively different results to the MFA-parametrized lateral interaction model. Importantly, only the explicit site-resolving KMC results are able to reproduce the experimentally observed selectivity trends for large particles. The case of small particles is outside the scope of the present work as the low-index Rh(211) and Rh(111) surfaces mainly represent step and terrace sites at larger particles free from finite-size effects. 
    
\begin{figure}[t]
        \centering
        \includegraphics[width=0.9\columnwidth]{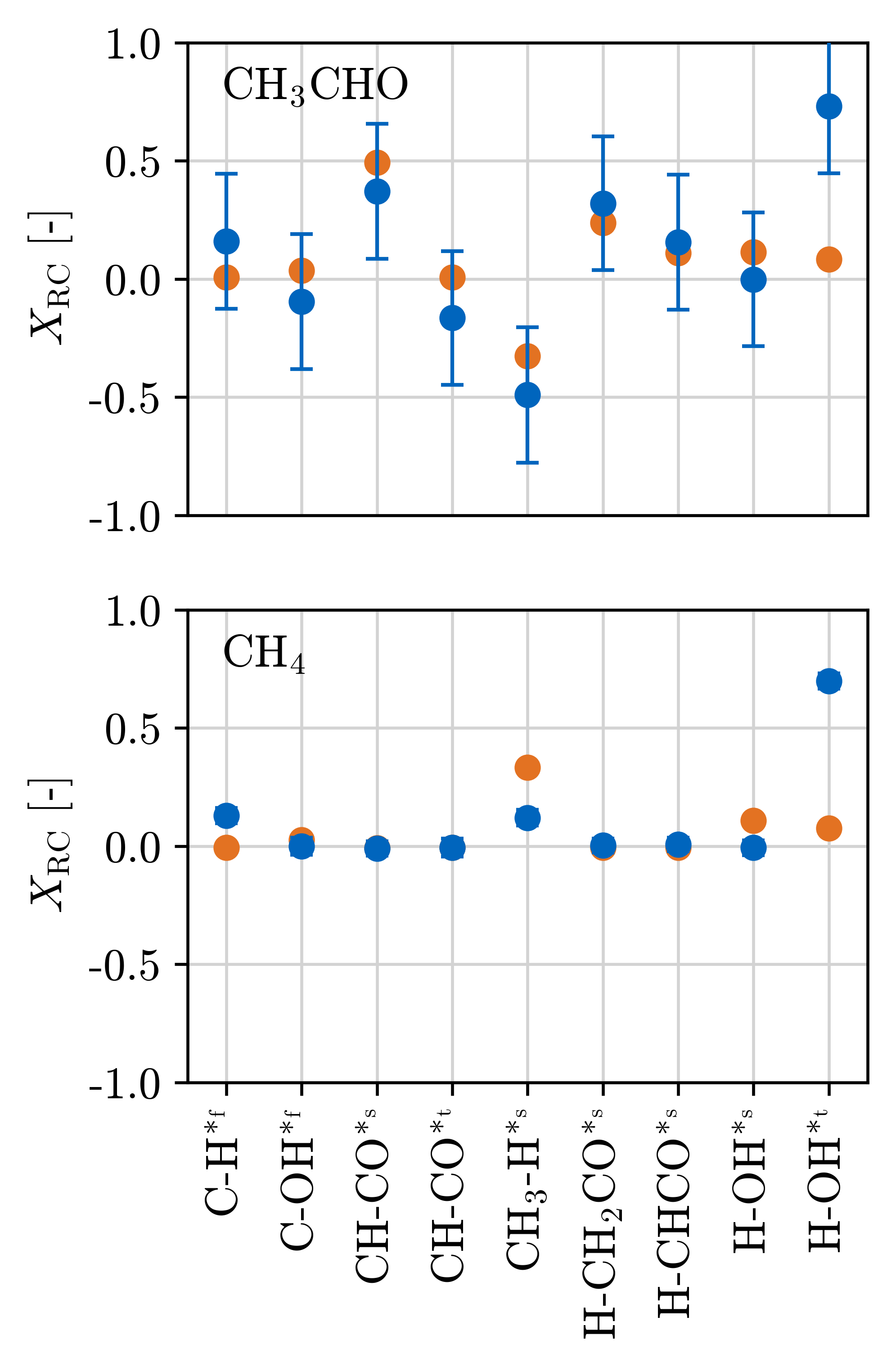}
        \caption{Sensitivity analysis of the KMC (blue points) and MFA (orange points) models for Rh(211) without lateral interactions, showing the DRC for $\mathrm{CH_{4,(g)}}$ (bottom) and $\mathrm{CH_{3}CHO_{(g)}}$ (top) of selected transition states. The analysis is performed at 650~K.
        }
        \label{fig:sensitivity}
    \end{figure}

    We believe that the shortcomings of the MFA model for both Rh(211) and Rh(111) are caused by the well-known problems of these models with accounting for effects of correlations and fluctuations, including fluctuations in the local coverage.\cite{Matera2011,Stamatakis2016,Liu2016,Li2021} In the following we exemplarily analyse this for the Rh(211) surface without interactions. This analysis starts by performing a sensitivity analysis to identify the rate-limiting steps (RLSs) of the two models. As shown in Figure~\ref{fig:sensitivity}, the RLS for both the methane and the acetaldehyde pathways is mainly water formation at the $t$ site in the KMC model, whereas for the MFA model it is mainly methane formation and CH-CO coupling at the $s$ site. The sensitivity analysis for the KMC model with interactions (see Figure~S3 in the SI) is very similar to that of the KMC model without interactions.

\begin{figure}
        \centering
        \includegraphics[width=\columnwidth]{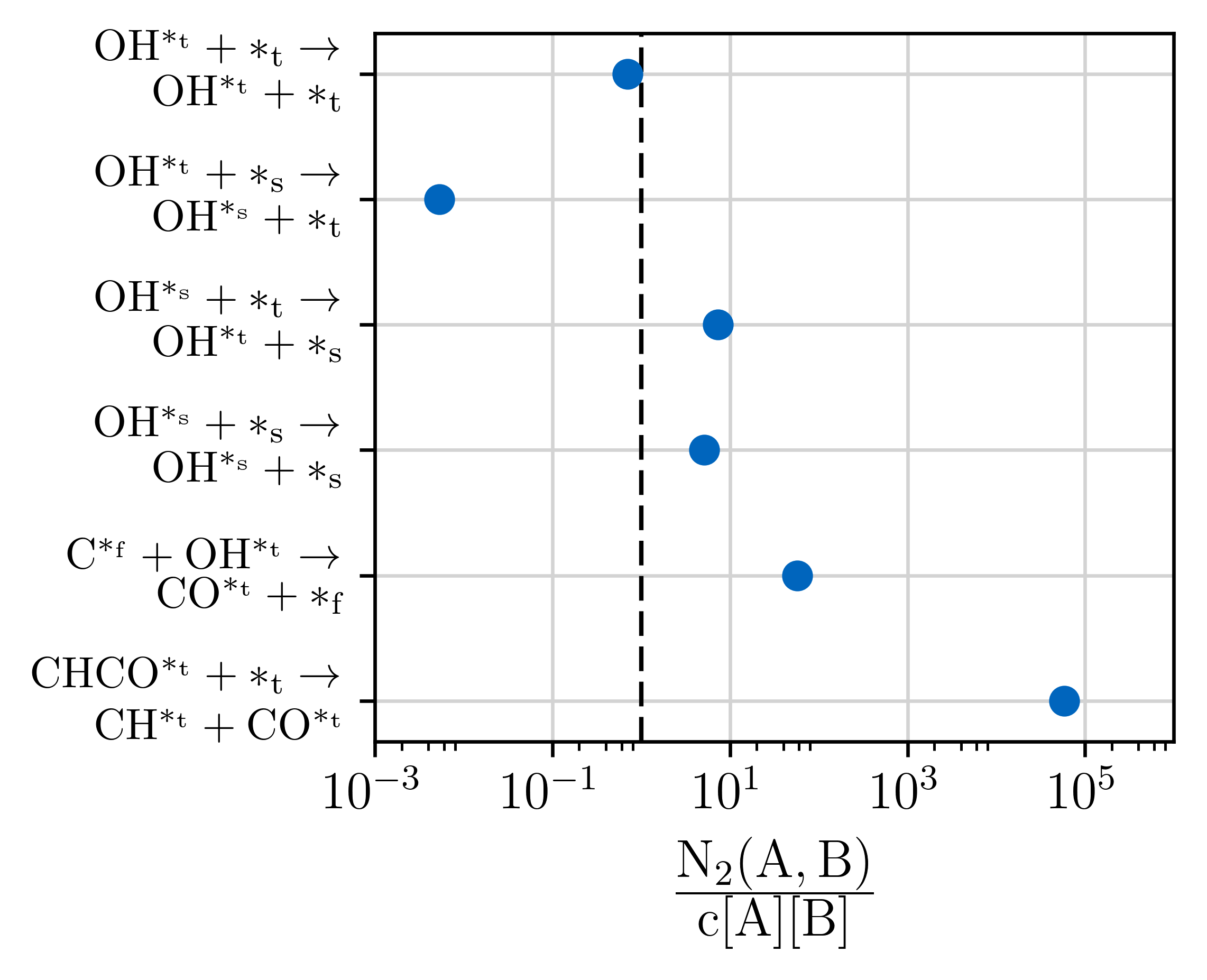}
        \caption{Ratio between KMC and MFA probabilities to find the pair of reacting species in selected second-order reaction steps at neighboring sites in the Rh(211) model without lateral interactions. The analysis is performed at 650~K.}
        \label{fig:pair_probability}
    \end{figure}

    Since a breakdown of the MFA is typically associated with second-order reaction steps where correlations in the spatial distribution of the two reacting species at the surface occur,\cite{Andersen2017akMC} we next evaluate the pair probabilities for selected elementary steps from our KMC simulations. In Figure~\ref{fig:pair_probability} we plot the ratio between the KMC-simulated pair probability to find the reacting species $A$ and $B$ at neighboring sites and the MFA-assumed probability equal to $c[A][B]$, where $c$ is a geometric factor (the site connectivity) and $[A]$ ($[B]$) is the surface coverage of species $A$ ($B$). It is seen that the MFA indeed breaks down for several of these steps, i.e.\ the ratio is significantly different from one. In particular, we note that the MFA breaks down for the reaction between C at the $f$ site and OH at the $t$ site to form CO at the $t$ site and H. This is the reverse of the CO dissociation step, which is executed just before the RLS in the KMC model where OH at the $t$ site is further hydrogenated to form water. The probability of finding the two reactant species, $\mathrm{C^{*_f}}$ and $\mathrm{OH^{*_t}}$, next to each other is about 57 times larger in KMC than in MFA. This is caused by the high barrier for C to diffuse along the step and the high CO coverage at the $t$ site, which effectively hinders OH diffusion at the $t$ sites, cf.\ our previous work on the comparison of MFA and KMC models for CO hydrogenation at stepped metals.\cite{Andersen2017akMC}

\begin{figure*}
        \centering
        \hfill%
        \subfloat{a)\addtocounter{subfigure}{-1}}
        \subfloat{%
        \includegraphics[valign=t,height=0.017\textwidth]{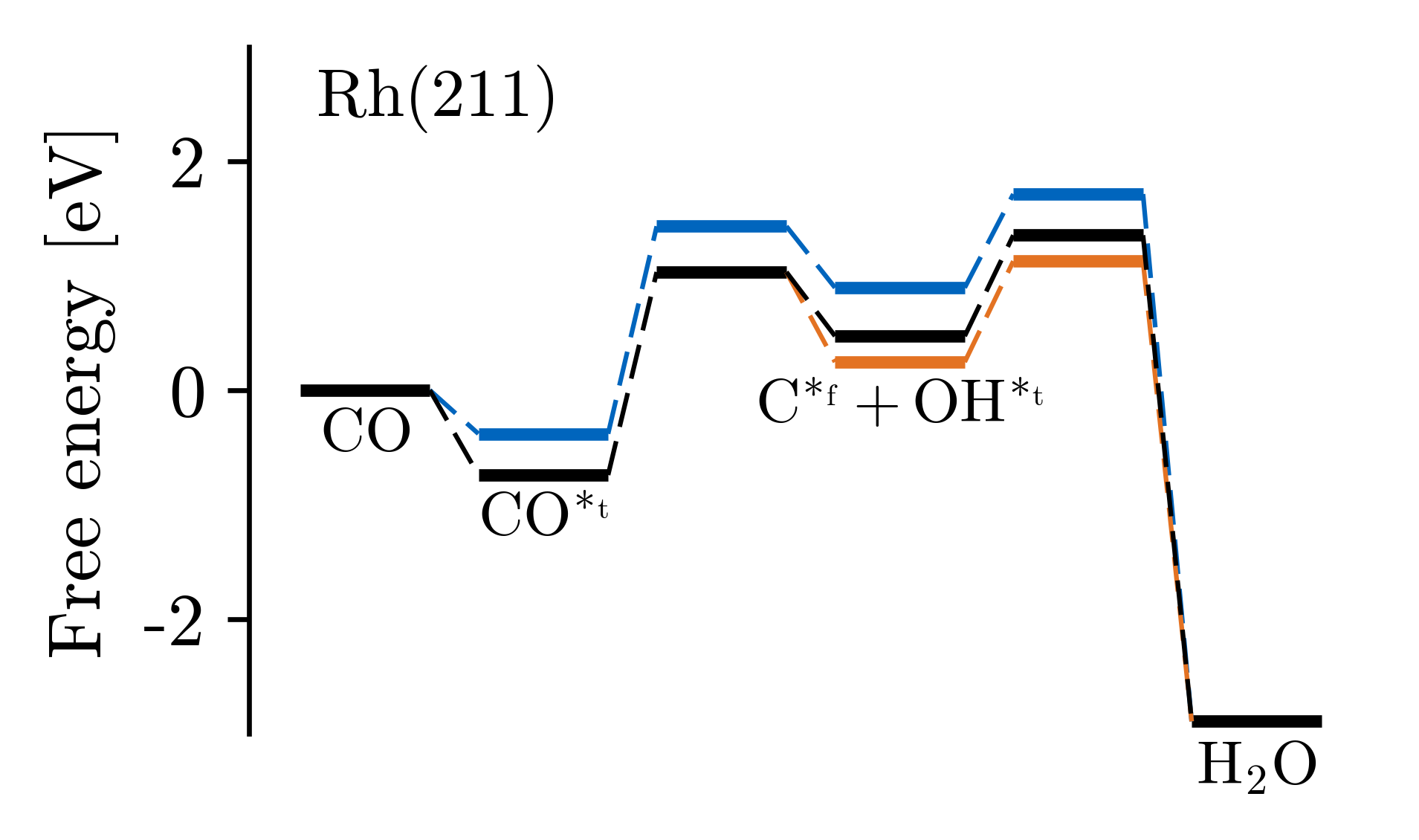}%
        \label{fig:PED_211}}%
        \hfill%
        \subfloat{b)\addtocounter{subfigure}{-1}}
        \subfloat{%
        \includegraphics[valign=t,height=0.017\textwidth]{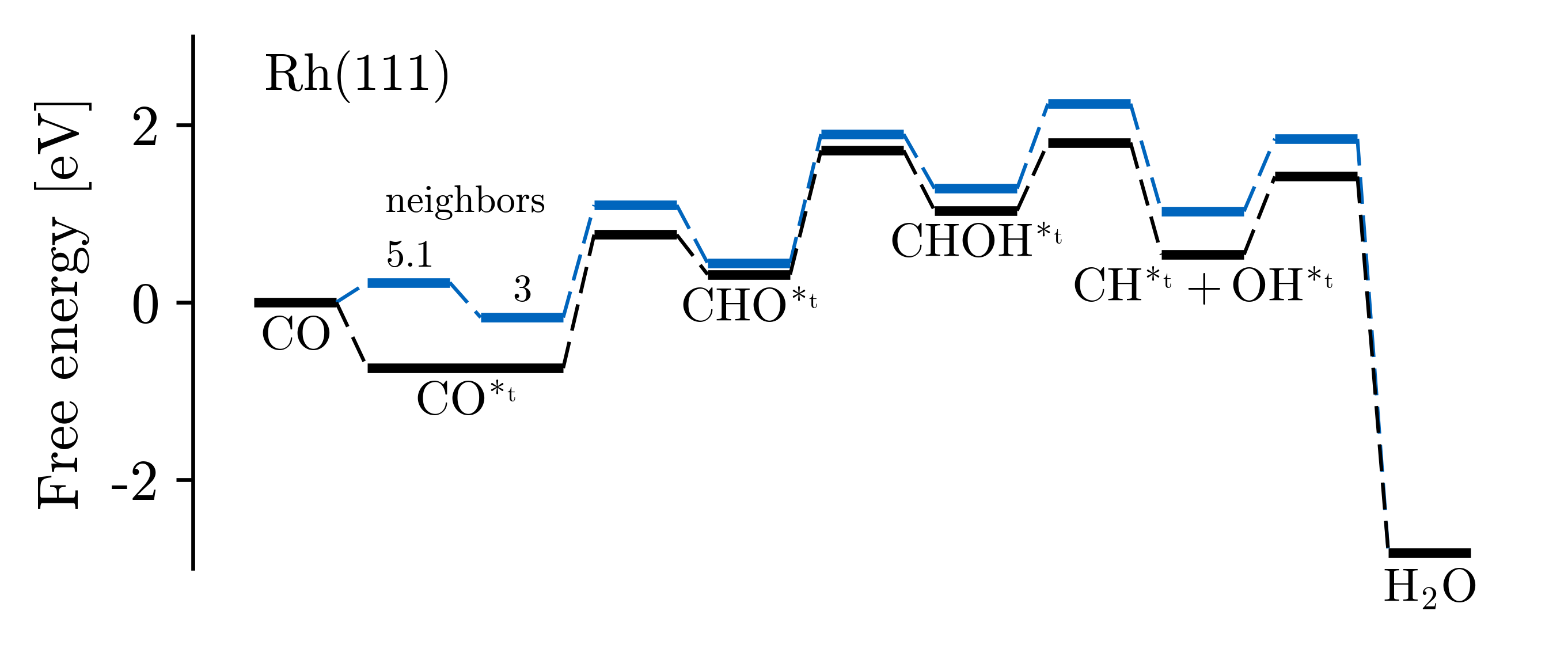}%
        \label{fig:PED_111}}%
        \hfill\phantom{}%
        \caption{Free energy diagrams for the CO dissociation and water formation pathways in the (a) Rh(211) and (b) Rh(111) KMC model without (black) and with (blue) lateral interactions. 
        In (a) the orange curve represents the modified probability for the reverse CO activation step in the MFA model without interactions (see text).
        The energy levels with lateral interactions for Rh(211) and CO desorption for Rh(111) are determined by the average barriers of the up to 10,000 last executed events of each process. For the remaining levels at Rh(111) we do not have enough statistics, e.g.\ the back reaction of CH and OH to form CHOH never occurs. Since these steps are mostly executed from the surface configuration illustrated in Figure S5 where each CO has three neighboring CO molecules, we instead determined the energy levels and barriers specifically for this surface configuration. The analyses are performed at 650~K.
        }
        \label{fig:PED}
    \end{figure*}

    As a consequence of this persistent correlation, it becomes more probable for $\mathrm{C^{*_{f}}}$ and $\mathrm{OH^{*_{t}}}$ to react back to form $\mathrm{CO^{*_{t}}}$ and H in the KMC model compared to the MFA model. In a free energy diagram, cf.\ orange and black lines in Figure~\ref{fig:PED_211}, this change in pair probability can be represented as an increase in the free energy barrier for the reverse reaction step of $\mathrm{C^{*_{f}}+OH^{*_{t}}}$ in the MFA model (orange line) caused by a decrease in the free energy of the $\mathrm{C^{*_{f}}+OH^{*_{t}}}$ state (i.e.\ a higher configurational entropy in the MFA-assumed well-mixed state). Since the free energy barrier for further reaction out of the $\mathrm{C^{*_{f}}+OH^{*_{t}}}$ state (i.e.\ water formation at the $t$ site) is unchanged, the effective barrier in the MFA free energy landscape (difference between $\mathrm{H\mhyphen OH^{*_{t}}}$ transition state and $\mathrm{CO^{*_{t}}}$ state) is smaller than the effective barrier in the KMC free energy landscape. This explains the larger TOF obtained in the MFA model and the change of the RLS from water formation at the $t$ site to methane formation and CH-CO coupling at the $s$ site (the latter steps are not shown in Figure~\ref{fig:PED_211}).

    From the free energy diagram in Figure~\ref{fig:PED_211} we can also explain why the KMC TOF and RLS for both methane and acetaldehyde formation at the Rh(211) facet does not change much upon inclusion of lateral interactions. As seen by comparing the black curve (without interactions) and the blue curve (with interactions), the free energies of the states that determine the effective barrier ($\mathrm{H\mhyphen OH^{*_t}}$ transition state and $\mathrm{CO^{*_t}}$ state) move up in energy by a similar amount as a consequence of interactions, causing the effective barrier and TOF to remain at a similar value.
    The detailed lateral interaction parameters at play in the two states are illustrated in Figure~S4 in the SI.
    
    While unimportant for the Rh(211) model, we already mentioned above that lateral interactions do have a huge effect on the Rh(111) model, and we will next analyze why this is the case. The free energy diagram with and without lateral interactions for the Rh(111) facet is shown in Figure~\ref{fig:PED_111}. The biggest difference compared to the Rh(211) facet is that here lateral interactions actually push the free energy of the $\mathrm{CO^{*_t}}$ state up close to the free energy of gas-phase CO. Thereby, the average CO coverage decreases from around 100\% without interactions to about 62\% with interactions at the analyzed temperature of 650 K, cf.\ Section S3.3 in the SI. In Figure~\ref{fig:PED_111} the $\mathrm{CO^{*_t}}$ state is split into two levels since CO desorption primarily happens from a local high-coverage state where on average 5.1 out of 6 neighboring sites are occupied by CO, whereas hydrogenation of CO primarily happens from a lower-coverage state where only 3 neighboring sites are occupied by CO. The average coverage pattern with about $\nicefrac{2}{3}$ CO coverage resembles a honeycomb lattice of CO molecules as illustrated in Figure S5 in the SI. The RLS for both methane and acetaldehyde formation in the model with interactions is primarily dissociation of CHOH to form CH and OH (and for acetaldehyde also the CH$_2$CO hydrogenation step), cf.\ Figure S6 in the SI.
    
    For Rh(111), the profound impact of lateral interactions on the simulation results cannot be explained by a change to the effective free energy barrier (i.e.\ the difference between the rate-limiting $\mathrm{CH\mhyphen OH^{*_t}}$ TS and the $\mathrm{CO^{*_t}}$ state in Figure~\ref{fig:PED_111}), since this barrier decreases with only around 132~meV upon inclusion of lateral interactions. Rather, the important point is that for CHOH to dissociate, a neighboring $t$ vacancy is required. Without lateral interactions, the surface is completely poisoned with an average CO coverage close to 100\%. Thus, the probability for CHOH dissociation to occur is negligible, which explains why no product formation could be observed in the KMC simulations. With interactions, however, a formed CHOH will on average have 3 CO and 3 vacancy neighbors (see Figure S5 in the SI), which makes the dissociation reaction possible. For the Rh(211) facet a similar CO poisoning is not observed, since here CO dissociation relies on a vacant $f$ site, and the vacancy coverage of the $f$ sites remains high even without lateral interactions, cf.\ Table S4.

    Up until now, we have shown that the MFA can break down both in the absence of lateral interactions (analyzed for the Rh(211) facet where the TOFs are overestimated in the MFA model compared to KMC) and in the presence of lateral interactions (analyzed for the Rh(111) facet where wrong selectivity trends are obtained in the MFA model compared to experiments). However, even if the KMC model does recover the experimental selectivity trends, the absolute TOF predicted by both the KMC model and the MFA model is not in good agreement with the experiments. 
    
    In Figure~\ref{fig:TOF_comparison_211} and Figure~\ref{fig:TOF_comparison_111} we mark with a blue diamond the total TOF (primarily methane) measured for the largest Rh nanoparticles investigated by Schumann~\textit{et~al.}\cite{Schumann2021} It is larger by about five orders of magnitude compared to the Rh(211) KMC simulation with lateral interactions (cf.\ Figure~\ref{fig:TOF_comparison_211}) and by about 5-6 orders of magnitude for the Rh(111) KMC and MFA simulations with lateral interactions (cf.\ Figure~\ref{fig:TOF_comparison_111}). The consistent, gross underestimation of the TOF by both KMC and MFA and for both investigated facets leads us to suspect that this discrepancy has its origin in the DFT calculations used to parametrize the kinetic models. In particular, it is well-known that generalized gradient approximation (GGA) functionals tend to overbind CO on transition metals,\cite{Patra2019} e.g.\ the CO adsorption energy obtained with the here employed BEEF-vdW functional for Rh(111) ($-$1.7~eV) is by 0.25~eV stronger than the experimental value ($-$1.45~eV\cite{Wei1997,AbildPedersen2007}). Since the state with CO adsorbed at a terrace site directly influences the effective barrier of the reaction, cf.\ Figure~\ref{fig:PED}, this error directly influences the obtained TOF.

\begin{figure*}
        \centering
        \hfill%
        \subfloat{a)\addtocounter{subfigure}{-1}}
        \subfloat{%
        \includegraphics[valign=t,width=0.45\textwidth]{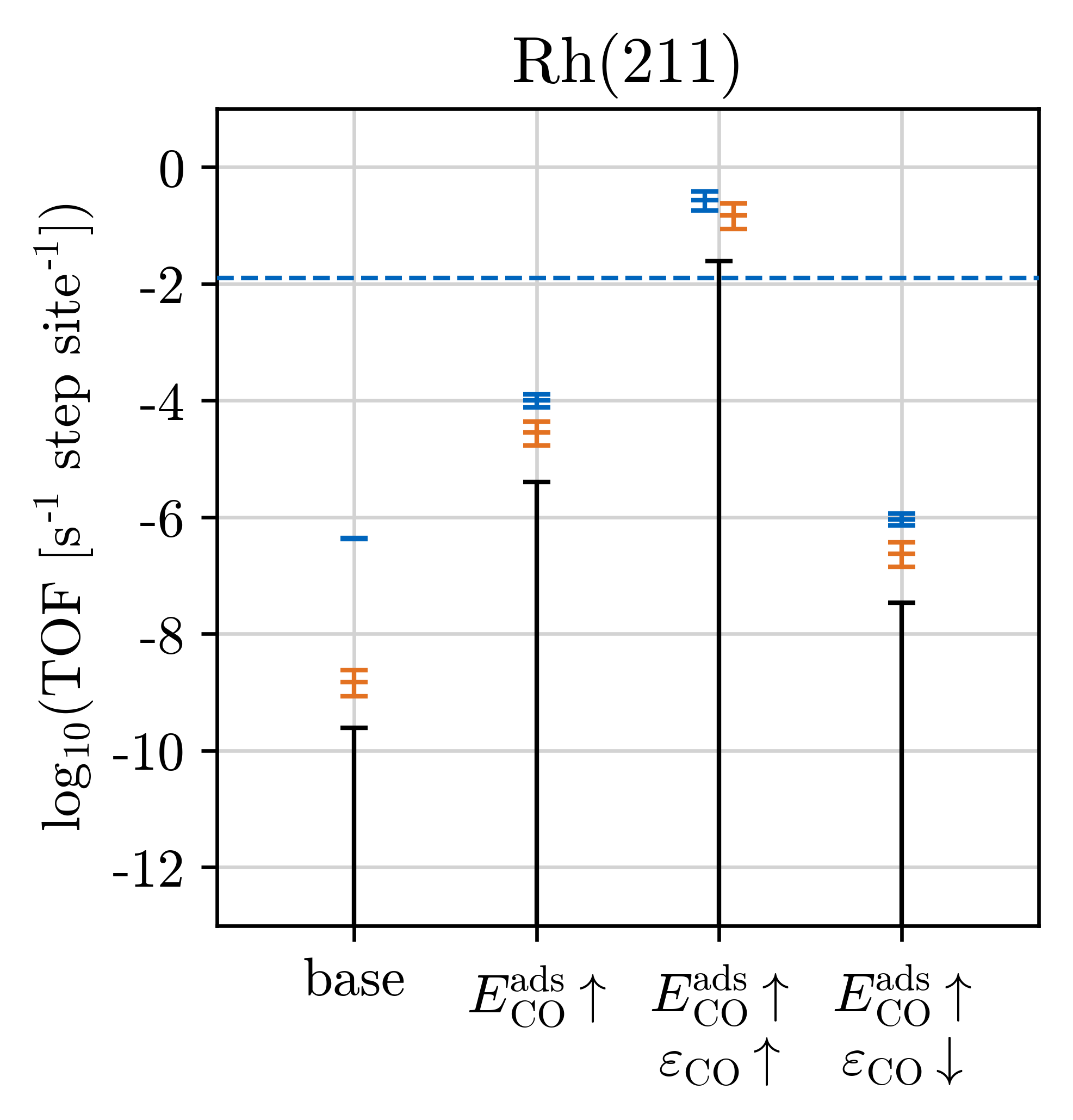}\label{fig:Eads_li_comparison_211}}%
        \hfill%
        \subfloat{b)\addtocounter{subfigure}{-1}}
        \subfloat{%
        \includegraphics[valign=t,width=0.45\textwidth]{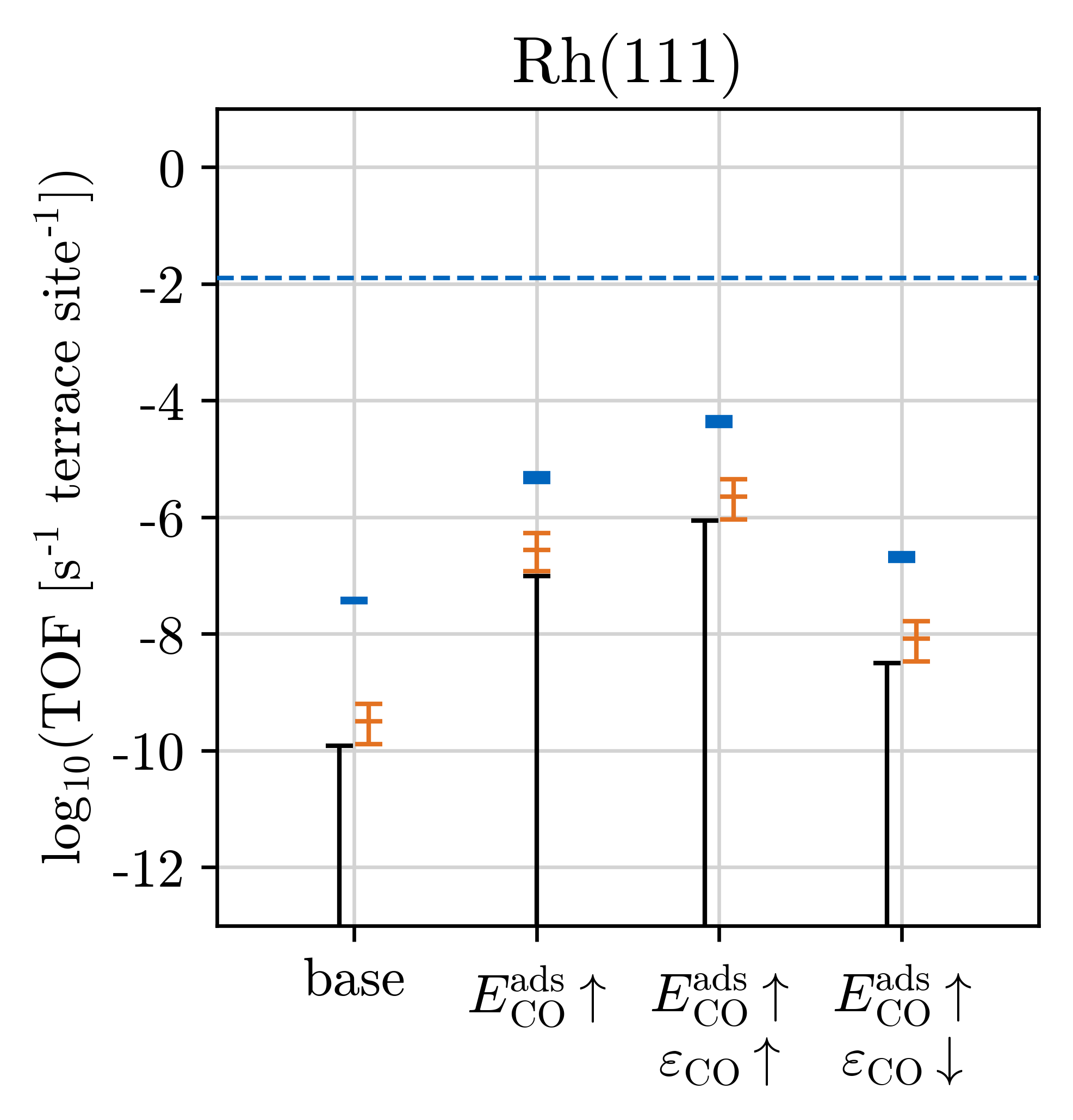}\label{fig:Eads_li_comparison_111}}%
        \hfill\phantom{}%
        \vskip0.1cm%
        \hfill%
        \subfloat{c)\hphantom{\ }\addtocounter{subfigure}{-1}}%
        \subfloat{%
        \includegraphics[valign=t,width=0.45\textwidth]{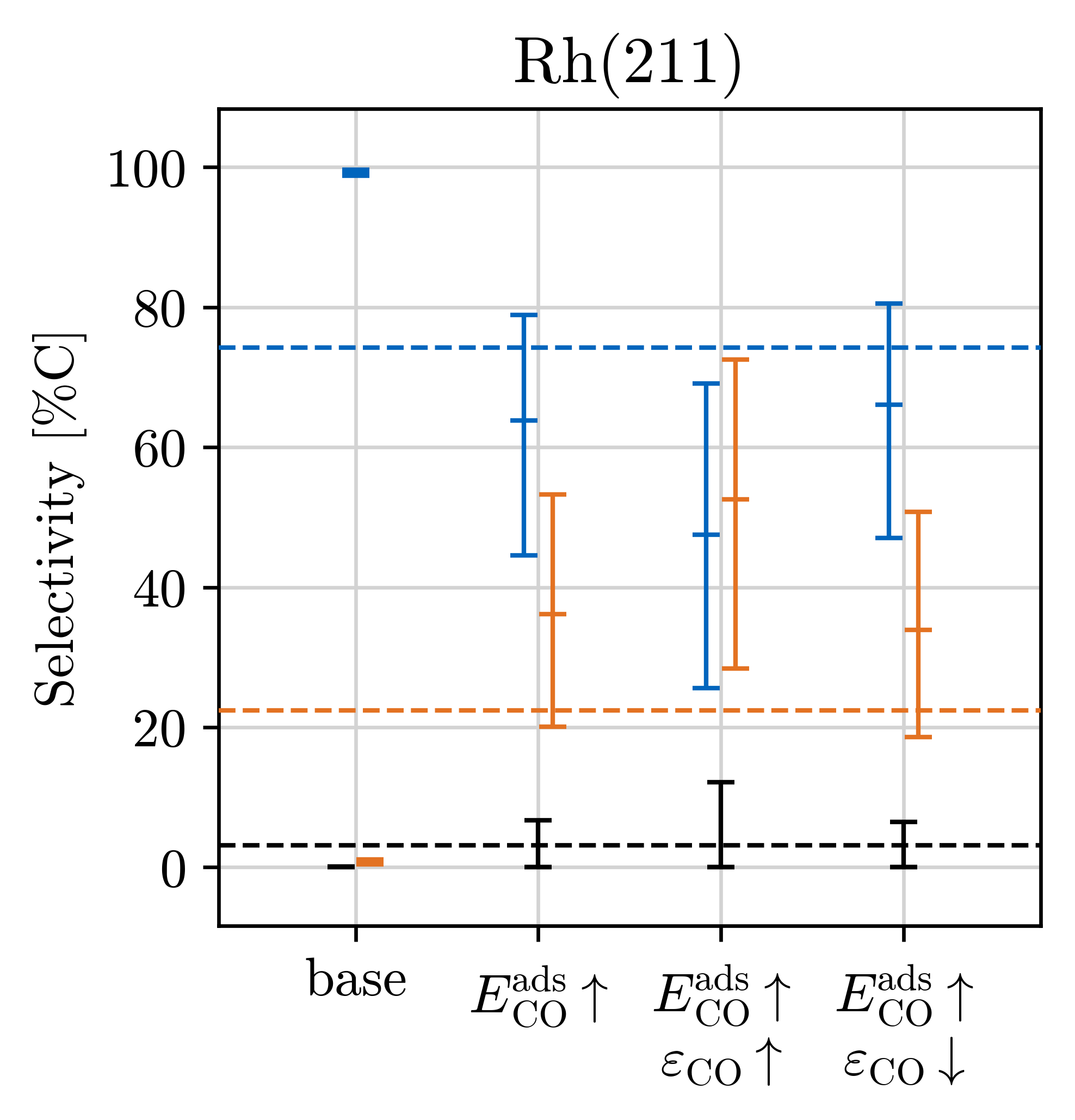}%
        \label{fig:Eads_li_comparison_selectivity_211}}%
        \hfill%
        \subfloat{d)\hphantom{\ }\addtocounter{subfigure}{-1}}%
        \subfloat{%
        \includegraphics[valign=t,width=0.45\textwidth]{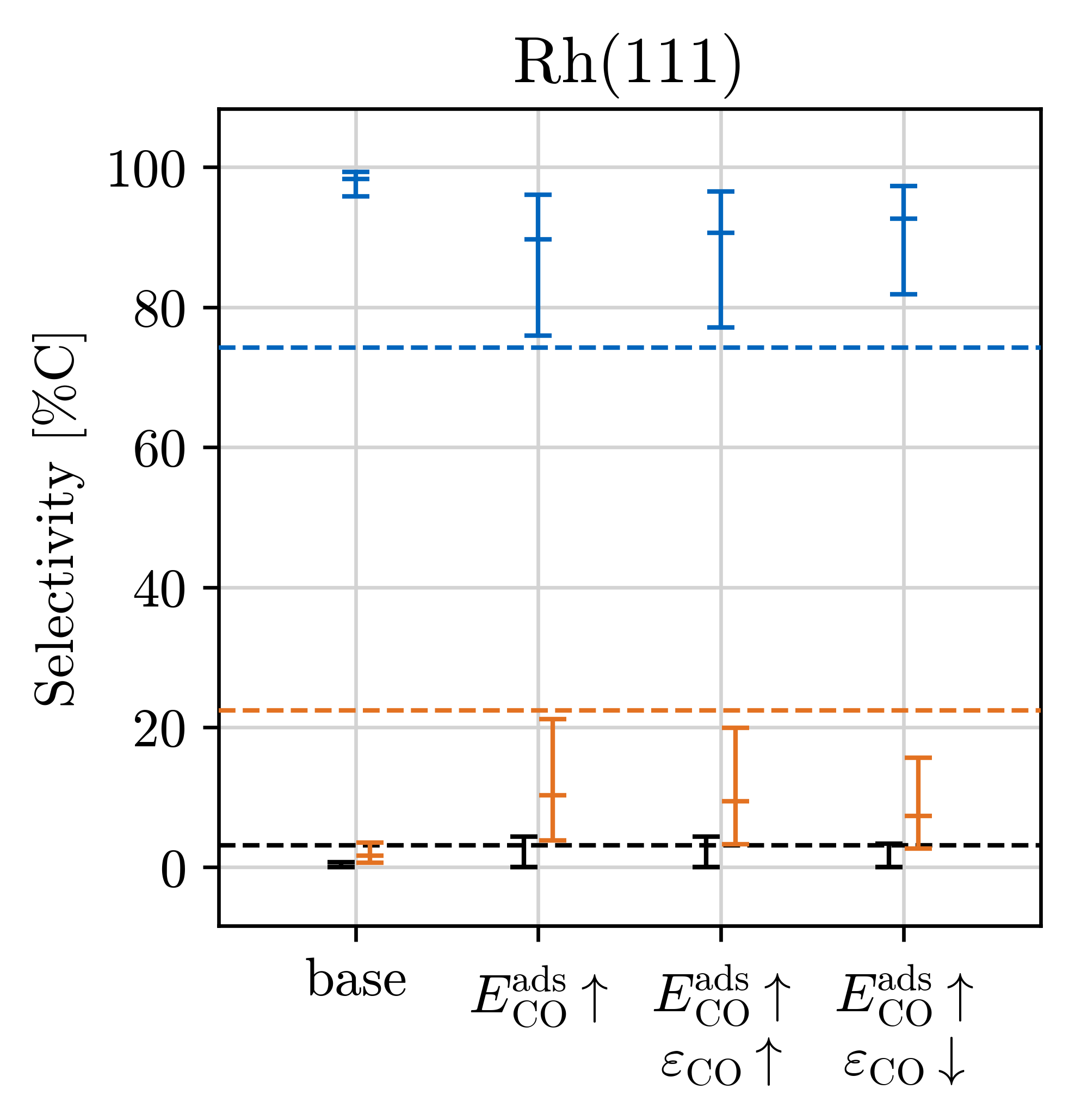}%
        \label{fig:Eads_li_comparison_selectivity_111}%
        }\hfill\phantom{}%
        \caption{Variations in predicted KMC TOFs for (a) the Rh(211) and (b) the Rh(111) facet and corresponding carbon selectivities for (c) the Rh(211) and (d) the Rh(111) facet with modifications to the DFT parameters in the models with lateral interactions (see text). Results are shown for CH$_4$ (blue), CH$_3$CHO (orange), and CH$_3$CH$_2$OH (black). "base" represents the unmodified results also shown in Figures~\ref{fig:TOF_comparison_211} and \ref{fig:TOF_comparison_111}. The dashed horizontal lines are the experimentally measured total TOF (primarily methane, (a) and (b)) and selectivities ((c) and (d)) of the largest nanoparticles with diameters above 5~nm from Schumann~\textit{et~al.}\cite{Schumann2021}. Both the theoretical and the experimental results are obtained at 523~K.
        }\label{fig:Eads_li_comparison}
    \end{figure*}

    In order to assess the possible implications of this error, we show in Figure \ref{fig:Eads_li_comparison} the TOFs and selectivities obtained for the Rh(211) and Rh(111) facets for both the hitherto discussed base model and a model where the CO adsorption energies are increased by 0.25~eV ($E_{\mathrm{CO}}^{\mathrm{ads}} \uparrow$). This already improves the agreement with experiments significantly, i.e.\ the TOF predicted for the Rh(211) facet (Rh(111) facet) is now only about two (three) orders of magnitude lower than the experiment. Of course, this analysis neglects the fact that there could also be errors associated with other rate-controlling parameters in our kinetic model, e.g.\ the transition state for water formation at the terrace site or lateral interaction parameters affecting either the latter transition state or the CO adsorption energy. Since we do not know in which direction such other errors might point, we show as an example in Figure~\ref{fig:Eads_li_comparison_211} and Figure~\ref{fig:Eads_li_comparison_111} also results where the interaction parameter for self-interactions between the CO molecules at $t$ and $s$ sites are increased ($\varepsilon_{\mathrm{CO}} \uparrow$) or decreased ($\varepsilon_{\mathrm{CO}} \downarrow$) by 50~meV from their base values. An error of this magnitude seems quite reasonable since we use an approximate CE that is terminated after pairwise nearest neighbor interactions.
    Taking these results as plausible "error bars" on the theoretical results, we can now reconcile theory with experiment to the extent that the experimental TOF lies within (admittedly rough) error bars on the theoretical values for the Rh(211) facet. In reality, the experimentally used nanoparticles of course contain both step and terrace sites, and the actual TOF measured might be the result of an interplay between reaction steps taking place at both site types. Such bifunctional effects have been previously discussed in the literature,\cite{Andersen2016,Andersen2017bifunctionalcatalysis} and have indeed recently been demonstrated theoretically for e.g.\ CO oxidation at Pt nanoparticles \cite{Jorgensen2018} and hydrogen evolution reaction at jagged Pt nanowires.\cite{Gu2021} The finding here that the TOF is largest, and in best agreement with experiment, on the Rh(211) facet is related to the facile hydrogen-assisted CO dissociation at the step sites exposed by this facet. However, the subsequent rate-controlling water formation step takes place at the terrace site, thus, both types of sites are required for a high activity. An analysis of how this would play out at realistic nanoparticle geometries will be an intriguing topic for future work.
    
    Comparing the experimental selectivities shown as dashed horizontal lines in Figures~\ref{fig:Eads_li_comparison_selectivity_211} and \ref{fig:Eads_li_comparison_selectivity_111} with the theoretical selectivities reveals that for both facets variations in the adsorption energies and interaction energies (as well as combinations) bring the selectivities closer to the experimental measurements. The observed trend of an increase in the acetaldehyde selectivity with a destabilization of adsorbed CO for both Rh(211) and Rh(111) is caused by a decrease of the effective barriers in the acetaldehyde pathways involving the RLSs of CH-CO coupling (for Rh(211) as shown in Figure~\ref{fig:sensitivity}) and CH$_2$CO hydrogenation (for Rh(111) as shown in Figure S6). Nevertheless, for all parameter modifications, our results robustly indicate that the Rh(111) facet is selective to methane, whereas the Rh(211) facet exhibits a similar selectivity to methane and acetaldehyde. Given that realistic 5 nm nanoparticles will dominantly expose the Rh(111) facet, as well as step and corner sites to a smaller extent, the sum of the theoretical selectivity trends should overall result in a methane selectivity in quite good agreement with the experimental measurements.
    
\section{Conclusions}

    We investigated the CO hydrogenation reaction on Rh(111) and Rh(211) using accelerated first-principles KMC simulations with and without lateral interactions parametrized from a cluster expansion model. The results are compared to MFA simulations from Yang~\textit{et~al.}\cite{Yang2016} with and without coverage dependence in the rate expressions. The coverage-dependent MFA model predicts acetaldehyde selectivity for the Rh(111) facet below $\sim$580~K, which is however in contrast to the KMC simulations with lateral interactions that predict methane selectivity at all temperatures. Importantly, only the KMC results are in agreement with recent detailed experimental investigations on selectivity trends of Rh nanoparticles from Schumann~\textit{et~al.}\cite{Schumann2021} 
    
    The inclusion of lateral interactions are found to have a huge effect on the Rh(111) simulations. We explain this from the fact that the RLS in this model (CHOH dissociation) requires a neighboring vacant terrace site, which is a probable lattice configuration only when taking into account the reduction of the surface CO coverage caused by repulsive CO-CO interactions. In contrast, for the Rh(211) model we find that the TOF and RLS (water formation at the terrace site) are only weakly affected by lateral interactions. This is explained from similar surface coverages with and without lateral interactions and from the fact that interactions shift the two states controlling the effective barrier in the dominant reaction pathway by similar amounts, leading to net similar effective barriers with and without interactions. For Rh(211) we furthermore show that the MFA breaks down also in the absence of interactions due to diffusion limitations and reaction-induced lattice inhomogeneities. This leads to higher TOFs and changes in the RLSs in the MFA model compared to the KMC model.
    
    Finally, we compare the overall activities (methane and acetaldehyde TOFs) to the experimental measurements from Schumann~\textit{et~al.} We find that we need to correct for the well-known CO overbinding of GGA functionals (about 0.25~eV for the here applied BEEF-vdW functional) in order to approach agreement with the experiments. This correction also improves the quantitative agreement with the carbon selectivites obtained by Schumann~\textit{et~al.} In particular, it leads to a larger increase in the acetaldehyde TOFs compared to the methane TOFs for both Rh(211) and Rh(111) since the effective barrier of the acetaldehyde pathway involves CO both in the CO activation step leading to CH$_x$ species at the surface (i.e.\ the step shared with the methane pathway) and in the subsequent step where the formed CH species reacts with another CO. 
    
    Overall, the insights obtained in this work could be relevant for further tailoring heterogeneous catalysts to improve their selectivity towards the desired higher oxygenates acetaldehyde and ethanol. The methodological advances demonstrated here, i.e.\ the combination of acceleration algorithms in KMC with efficient modeling of lateral interactions, open up possibilities for treating also complex reaction networks at a level of detail beyond the hitherto applied approximate MFA models. This is important in order to achieve reliable mechanistic insights as a solid basis for the rational design of selective catalysts.

\begin{suppinfo}
    
    Supporting Information. Additional details on the compared microkinetic models, DFT, and KMC. This material is available free of charge via the internet at http://pubs.acs.org. Input files for the kmos simulations for the different models together with additionally calculated DFT geometries are available at https://github.com/m-deimel/CO\_hydrogenation.git.
    
\end{suppinfo}

\begin{acknowledgement}

    We acknowledge funding and support from the Deutsche Forschungsgemeinschaft (DFG, German Research Foundation) under Germany’s Excellence Strategy - EXC 2089/1- 390776260. M.A.\ acknowledges funding from the European Union’s Horizon 2020 research and innovation programme under the Marie Sk\l{}odowska-Curie grant agreement No 754513, the Aarhus University Research Foundation, the Danish National Research Foundation through the Center of Excellence 'InterCat' (Grant agreement no.: DNRF150) and VILLUM FONDEN (grant no.\ 37381).
    
\end{acknowledgement}

\bibliography{references.bib}

\end{document}